\documentclass[usenatbib,usedcolumn]{mnras}
\usepackage{graphicx}
\usepackage{amsmath}
\usepackage[dvipsnames]{xcolor}
\usepackage{bm}
\usepackage{natbib}
\usepackage{amssymb}
\usepackage{newtxtext}
\usepackage[varvw]{newtxmath}
\usepackage{threeparttable}
\usepackage{ulem}
\usepackage{soul}
\usepackage{tablefootnote}
\definecolor{darkgreen}{rgb}{0.15,0.5,0.15}
\definecolor{darkblue}{rgb}{0.15,0.15,0.5}

\newcommand{\kpc}{{\rm kpc}}

\title[]{Scalar Field Dark Matter: Impact of Supernovae-driven blowouts on the soliton structure of low mass dark matter halos}

\author[Victor H. Robles, J. L. Zagorac and N. Padmanabhan]{
Victor H. Robles,$^{1,2,3}$\thanks{E-mail:roblev@rpi.edu}
J. Luna Zagorac,$^{3,4}$
and Nikhil Padmanabhan$^{3}$
\\
$1$ Department of Physics, Applied Physics and Astronomy, Rensselaer Polytechnic Institute, Troy, NY, 12180, USA\\
$2$ Yale Center for Astronomy and Astrophysics, Yale University, New Haven, CT 06520, USA\\
$3$ Department of Physics, Yale University, New Haven, CT 06520, USA \\
$4$ Perimeter Institute for Theoretical Physics, 31 Caroline St. N, Waterloo, ON N2L2Y5, Canada
       }

\begin{document}
\date{MNRAS, xxx }

\pagerange{\pageref{firstpage}--\pageref{lastpage}} \pubyear{2018}

\maketitle

\label{firstpage}

\begin{abstract}
We present the first study on the gravitational impact of supernova feedback in an isolated soliton and a spherically symmetric dwarf SFDM halo of virial mass $1\times 10^{10}\mathrm{M_\odot}$. We use a boson mass $m=10^{-22}\mathrm{eV/c^2}$ and a soliton core $r_c \approx 0.7$kpc, comparable to typical half-light radii of Local Group dwarf galaxies. We simulate the rapid gas removal from the center of the soliton by a concentric external time-dependent Hernquist potential. We explore two scenarios of feedback blowouts: \textit{i}) a massive single burst, and ii) multiple consecutive blowouts injecting the same total energy to the system, including various magnitudes for the blowouts in both scenarios. In all cases, we find one single blowout has a stronger effect on reducing the soliton central density. Feedback leads to central soliton densities that oscillate quasi-periodically for an isolated soliton and stochastically for a SFDM halo. The range in the density amplitude depends on the strength of the blowout, however we observe typical variations of a factor of $\geqslant$2. One important consequence of the stochastic fluctuating densities is that, if we had no prior knowledge of the system evolution, we can only know the configuration profile at a specific time within some accuracy. 
By fitting soliton profiles at different times to our simulated structures, we found the (1-$\sigma$) scatter of their time-dependent density profiles. For configurations within the 1$\sigma$ range, we find the inferred boson mass 
is typically less than 20\% different from the real value used in our simulations. Finally, we compare the observed dynamical masses of field dwarf galaxies in our Local Group with the implied range of viable solitons from our simulations and find good agreement.
\end{abstract}

\section{Introduction}\label{sec:intro}

The standard Cold Dark Matter (CDM) paradigm is in remarkable agreement with the
large-scale structure of the Universe. At small scales ($\sim 1$ \kpc), however,
this paradigm faces some discrepancies with the observationally inferred
properties of low mass galaxies (see \cite{bullock17} for a review), most
notably for the dwarf galaxies within our local neighbourhood. While the stellar
luminosites of the nearby dwarf galaxies in our Galaxy span three orders of
magnitude, their inferred gravitational mass is consistent with a host dark
matter halo of virial mass $M_h \approx 10^{10} \mathrm{M_\odot}$, making these low mass
systems highly dark-matter dominated and ideal candidates to test the
predictions for their dark matter distributions. These challenges include the cusp-core problem,
where the inferred central densities of some dwarf galaxies (of typical halo mass
$\sim 10^{10} \mathrm{M_\odot}$) are less dense than what is predicted from the
dark-matter-only (DMO) CDM simulations. Another challenge is the
too-big-to-fail (TBTF) issue, which highlights that DMO CDM cosmological
simulations of a Milky Way mass halo predict a large abundance of massive
satellite halos inconsistent with the lower
number of the brightest (most massive) dwarf galaxies that are currently
observed \citep{boylan11}.

Numerous works have attempted to resolve the cusp-core issue.
For instance, it is argued that some caveats of using observations of the
gas rotation and stellar kinematics in dwarf galaxies to infer the existence of
a shallow inner density (core) \citep{moore1994,wp11,deBlok08,deNaray08} are
underestimating uncertainties in the non-circular motions of the gas component
that depend on the observer's line-of-sight \citep{oman19,santos20} and
neglecting gravitational effects that could arise if the underlying potential is
not spherically symmetric \citep{genina17}. To date, it remains plausible that
cores lie within these (dark matter dominated) dwarf galaxies. This puzzle has
led to two main scenarios that could explain the shallow densities : i) the
central density cores are the result of new dark matter physics (different from
the standard CDM model) that manifest at the scale of dwarf galaxies, ii) core
formation emerges from a non-negligible gravitational impact that the baryonic
matter has on the dark matter distribution, via various astrophysical processes
(e.g. supernovae feedback) \citep{navarro96,governato12,chan15,fitts17}.

One  alternative dark matter model that has recently gained interest due to its potential to solve the small scale issues is the Scalar Field Dark Matter (SFDM) model (also known as Bose-Einstein Condensate DM, axion-like dark matter, Fuzzy/Wave DM, and UltraLight DM). This model assumes the dark matter is composed of spin-0 particles of mass of $m \sim 10^{-22} \mathrm{eV/c^2}$ that decouple non-thermally from the rest of the matter in the early universe \citep{lee96,matos00,hu00,suarez14,hui17,urena19}. This ultra-light mass implies a de Broglie wavelength $\lambda_{\rm dB} \sim $ 1 kpc, comparable to the half-light radii of Milky Way dwarf spheroidal galaxies. This type of dark matter has wave-like features inside dark matter halos. Due to the Heisenberg uncertainty principle acting at the galactic scale the dark matter distribution at the center of galaxies can not collapse indefinitely, resulting in the formation of a central core. Many works have confirmed that the ground state self-gravitating SFDM configurations have a characteristic density profile called a \textit{soliton}. Solitons display a constant central density and are stable solutions under small perturbations \citep{guzman04}. These solitons are found to live at the center of dark matter halos in  cosmological SFDM simulations \citep{sch14}. These small scale features, added to the fact that at cosmological scales SFDM behaves as CDM, have increased the interest on SFDM as a viable alternative model that provides a natural explanation to the low central densities of dwarf galaxies \citep{Strigari08,rob12,chen17} and help to alleviate the Too-Big-to-Fail issue \citep{rob18}. 

Previous analysis in the SFDM model using the observational constraints from
dwarf galaxies suggest a preferred boson mass of $m_{22} = m/(10^{-22} \mathrm{eV/c^2})
\sim 1$ \citep{chen17,rob18,Lora15}. While typical low mass dwarf galaxies  have
half-light radius within $<1$kpc, for galaxies with much larger sizes we require
knowledge of the soliton and its outer dark matter halo. Cosmological
simulations in the SFDM paradigm reveal that the interplay of a soliton with its
surrounding dark matter halo leads to a relation between the inner soliton core
and its dark matter halo mass, albeit with large scatter
\citep{sch14,May21,chan22,2023PhRvD.107h3513Z}. \cite{bar18} fit the rotation curves of spiral
galaxies and prefer a higher boson mass. However,
\cite{kendall20} found that a boson mass $m_{22}\sim 1$ is allowed when the
scatter in the soliton core-halo mass relation is considered. On the other hand,
cosmological constraints coming from Lyman-$\alpha$ forest power spectrum from
high-redshift quasars seem to suggest a higher boson mass
\citep{irsic17,Zhang18,Nori18}. In the absence of self-consistent cosmological
hydrodynamics simulations in the SFDM model, it remains unclear how strongly
non-linear effects between the solitons and the baryonic matter at high redshift
would affect the cosmological constraints. 

Alternatively, rather than assuming new dark matter physics, numerous
hydrodynamics simulations within the CDM framework have invoked impulsive gas
outflows from the center of the halos originating from starbursts (supernovae
feedback) as a mechanism to transfer (gravitational) energy to the DM particles
\citep{governato12,fitts17,tollet16}, the cumulative impact of such episodic gas
blowouts eventually decreases the central DM density and a core could be formed
\citep{dicintio14,tollet16,chan15, fitts17, lazar20}. While core formation
driven by supernovae feedback remains debated, there seems to be a
consensus that this mechanism depends on the stellar-to-halo mass ratio of the
simulated halo \citep{fitts17,tollet16}. Since the most efficient scale is in the
range of bright dwarf galaxies (stellar mass $M_\star \sim 10^{6-7}\mathrm{M_\odot}$)
\citep{lazar20}, it remains plausible that the existence of a core in the
massive dwarf scale is sensitive to some details of the star formation algorithm
used in a hydrodynamics simulation \citep{Bose19,Benitez19,Dutton20}. For
galaxies whose stellar-to-halo masses are lower than bright dwarf galaxies, e.g.
faint and ultra faint dwarf galaxies, current CDM hydrodynamics simulations seem
to show more robustly a conspicuous cusp at the center of their dark matter
halos \citep{wheeler19,lazar20}. 

While in full hydrodynamics simulations it is possible to fully evolve the
non-linear interplay between baryons and dark matter as the galaxy evolves, the
complex astrophysical processes make it difficult to interpret which specific
process is responsible for core formation. This is further complicated when we account
for the  uncertainties in the star formation implementation (e.g. the star
formation threshold, how SNe energy and momentum is coupled to the gas).
However, since the dark matter will ultimately react to the gravitational
effects of the baryonic matter, we can attempt to single out the most energetic
astrophysical processes that (likely) lead to the strongest impact on the dark
matter density, one of them being the episodic outflows from supernovae. 

There have been some attempts to use analytic potentials to study supernovae
(SNe) blowouts in dwarf galaxies hosted by CDM dark matter halos
\citep{shea13,burger19,Freundlich19}. In \cite{shea13}, the authors introduced an
effective model of SNe feedback by adding a time-varying external potential at
the center of an non-cosmological dark matter dwarf halo. The potential mimicks
the gravitational effect of gas being accreted at center and the subsequent gas
outflows due to SNe feedback. These works found that for energetically plausible
SNe blowout events in the classical Milky Way dwarf spheroidal galaxies, the
central dark matter density is reduced within 1 kpc but a cusp profile still
remains, consistent with the results from CDM cosmological simulations at this
galaxy mass scale \citep{lazar20}. 

These idealized simulations confirm that even if baryons may be a small fraction
of the total mass in dwarf galaxies, they can have a observable impact in the
behavior of the dark matter. It is then necessary to study how the central dark
matter in alternative DM models respond to similar baryonic processes and how
their predictions are modified when baryons are neglected. In this regard, there
are only a few works that perform self-consistent hydrodynamics simulations of
dwarf galaxies beyond the CDM paradigm, including self-interacting dark matter
\citep{rob17,fry15,Vogelsberger14}, warm dark matter
\citep{Fitts19,Bozek18,Colin15}, and SFDM \citep{mocz19,mocz20,Veltmaat20}.
In this work we follow \cite{shea13} to provide a
more transparent approach, we focus on studying for the first time the
gravitational impact of SNe feedback in a soliton core whose size and mass are
consistent with current constraints from classical dwarf galaxies.

This manuscript is organized as follows. In Section 2 we provide the details of our  SFDM simulations and the model for the supernovae blowouts. In Section 3 we describe and discuss the results of our numerical runs, as well as compare with observations. In Section 4 we discuss our conclusions.

\section{Simulations}

\subsection{Initial conditions for self-gravitating solutions in SFDM }

In the non-relativistic regime, the scalar field dark matter evolution of a
self-gravitating configuration is described by a macroscopic wave function whose
dynamics are governed by the Schr\"odinger-Poisson (SP) system of equations. In
a non-expanding universe, they take the form:
\begin{equation}
i  \hbar \frac{\partial \psi}{\partial t} = \frac{-\hbar^2}{2m}\nabla^2 \psi + m V \psi,
\end{equation}

\begin{equation}
\nabla^2 V = 4\pi G m(|\psi|^2-  \langle |\psi|^2 \rangle),
\end{equation}
where $m$ is scalar field particle mass, $\psi$ is the wave function that is normarlized such that $\rho= m|\psi|^2$ is the dark matter mass density and $\bar{\rho}= m \langle |\psi|^2 \rangle$ is the mean mass density, $V$ is the  gravitational potential. In such a system, all particles share a common wave function, hence the mass density of the dark matter fluid traces the probability density distribution $|\psi|^2$.

The SP system remains invariant under a scaling symmetry with scaling parameter $\lambda$ as follows \citep{ji94}
\begin{equation}
\{t,x,V, \psi,\rho \} \rightarrow \{ \lambda^{-2} \hat{t},\lambda^{-1} \hat{x},\lambda^{2} \hat{V}, \lambda^{2} \hat{\psi}, \lambda^{4} \hat{\rho} \},
\label{lambdasym}
\end{equation}
additionally, given our definition of the dark matter density the system can also be scaled by the SFDM mass $m \rightarrow \alpha m$ \citep{mocz17} 
\begin{equation}
\{t,x,V, \psi,\rho \} \rightarrow \{ \alpha\hat{t}, \hat{x},\alpha^{-2}\hat{V},\alpha^{-3/2} \hat{\psi},\alpha^{-2} \hat{\rho} \}.
\label{masssym}
\end{equation}
Therefore, the resulting solutions from the SP system can be scaled to different boson masses. In this work, we will fix the boson mass $m_{22}=1$, this fiducial value is well-motivated  in the literature by its plausability to address the small scale prooblems in CDM, in particular at the dwarf galaxy scale.

There are different numerical methods to find the ground state solution (also called \textit{soliton}) to the SP system. While there is no analytic form for the soliton density profile, \cite{sch14} found that it can be well-parametrized by the following expression:
\begin{equation} 
\rho (r) =\frac{\rho_c}{\bigg( 1+0.091 \bigg( \frac{r}{r_c} \bigg) ^2\bigg )^8}
\label{rhosol}
\end{equation} 
where $\rho_c$ is the central soliton core density given by
\begin{equation}
\rho_c=1.93 \times 10^7 m_{22}^{-2}\bigg(\frac{r_c}{1 \rm kpc}\bigg)^{-4} \rm \mathrm{M_\odot} kpc^{-3},
\label{rhocore}
\end{equation}
and $r_c$ is the soliton core radius. Eq.~\eqref{rhosol} is accurate to 2\% to the numerical solution of an isolated soliton in equilibrium within $r\leq 3r_c$ and the soliton mass enclosed within this radius is $\approx$ 95\% of the total soliton mass. 

For an isolated soliton it is expected that its core size is similar to the de Broglie wavelength $r_c \sim \lambda_{ \mathrm{dB}}$. As solitons are the lowest energy solutions, they describe the smallest gravitationally bound structures that  form as the SFDM dark matter undergoes gravitational collapse. The subsequent growth depends on several non-linear processes, for instance, a dense environment around a soliton provides a mass reservoir from which the soliton could continue accreting mass and develop a large outer halo. On the other hand, a soliton living in a low density environment or experiencing tidal disruption could lead to structures whose total mass is mostly given by the soliton mass. 
Only a few cosmological SFDM simulations are available due to the highly demanding spatial and temporal resolution   needed to evolve large volumes until the present time to resolve the de Broglie wavelength in the halos. 
Using the central isolated halos in the cosmological SFDM simulations from \cite{sch14}, \cite{sch14b} obtained an empirical relation between the soliton core radius and its host halo virial mass $M_h$, at redshift $z=0$ it reads: 
\begin{equation}
r_c= 1.6 \,\mathrm{kpc} \bigg(\frac{M_h}{10^9 \mathrm{M_\odot}}\bigg)^{-1/3} m_{22}^{-1}
\label{rcMh}
\end{equation}
This soliton core-halo mass relation for isolated halos informs about the properties of the inner soliton to its outer dark matter halo and vice versa. Notably, a similar structural relation emerges in non-cosmological simulations in which solitons are formed either by undergoing gravitational collapse or by merging with other solitons \citep{mocz17,veltmaat18,sch14b}, in this closed systems the total mass of the simulation can be regarded as the halo mass. Motivated to capture the dynamics of the inner solitons hosting isolated dwarf galaxies, for our choice of boson mass $m_{22}=1$, current observations of the dynamical mass at the half-light radii of MW classical dwarf galaxies and in dwarf galaxies in the field constrain the soliton core sizes to be in the range $r_c=0.5-1$ kpc \citep{rob18,chen17}. From Eq. \eqref{rcMh} it implies these solitons would be hosted in halos of masses $M_{\rm h} \sim 10^{10} \mathrm{M_\odot}$.  

More recent work has put several caveats on this core-halo mass relation of SFDM: for example, \citep{2022MNRAS.511..943C, 2023PhRvD.107h3513Z} found that the size of the soliton of a given halo depends not only on the halo's total mass, but also on tidal stripping due to both the halo's physical interactions and numerical effects. The works cited within these two investigations reveal a variety of possibilities for the core-halo relationship in SFDM; indeed, \citep{2022PhRvD.105b3512Y} self-consistently construct halos with many relative core sizes using a wave superposition method. For concreteness, we will assume the structural relation presented in the previous paragraph, but acknowledge that the soliton core's size may depend on each halo's environment.  

Guided by these soliton size constraints ($r_c \lesssim 1 \, \rm{kpc}$), we explore how SNe feedback blowouts impact the halo center for two cases: an isolated soliton and a soliton that retains an non-zero outer halo component. For our SFDM dwarf halo we fix the soliton core $r_c=0.771$ kpc, Eq. \eqref{rhocore} implies that its soliton mass $M_{\mathrm{sol}}=2.968 \times 10^8 \mathrm{M_\odot}$. 
In Fig. 1, we show the density profile for an isolated soliton in equilibrium that has our fixed soliton core size (dashed line), and the density profile of a SFDM halo (solid line) with the same soliton core (mass) embedded in a virialized outer halo component. The SFDM halo was formed by merging solitons using the code {\sc chplUltra} \citep{chplUltraPaper}, and was then spherically- and time-averaged to obtain the smooth profile shown.

\subsection{Model for SNe blowouts}
\begin{figure}
    \centering
    \includegraphics[width=\columnwidth]{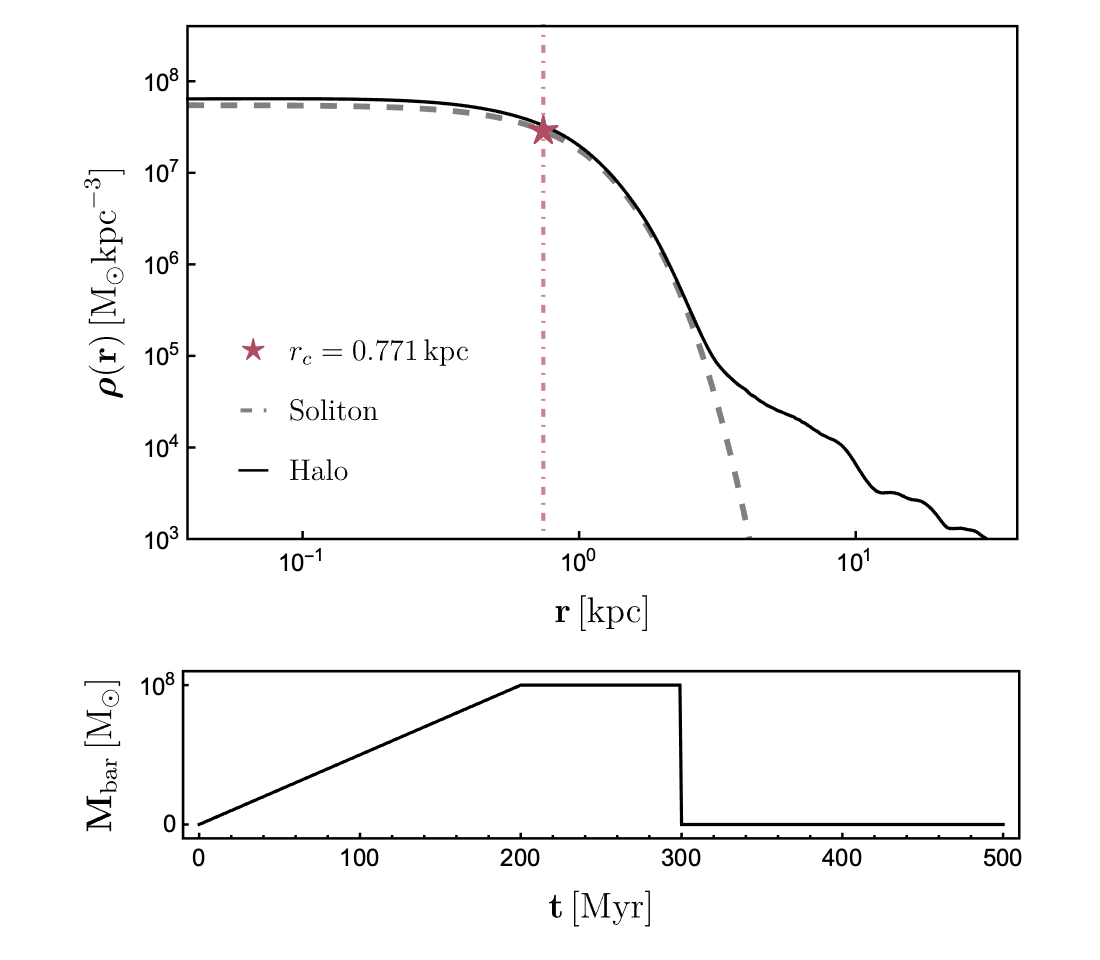}  
    \caption{Dark matter density profile of an isolated soliton in equilibrium
    with a core radius $r_c=0.771$ kpc (dashed). Also shown is a SFDM halo with the same initial soliton core radius (black line). Below, we show the evolution of the
    effective baryon mass $M_{\rm{bar}}(t)$ in our single-burst supernova model as described in the text.
    }
    \label{fig:fig1}
\end{figure}

To implement a simple model for the episode of gas accretion followed by a sudden gas blowout, we follow 
the procedure described in \cite{shea13}. We start by mimicking the gas accretion at the center of the dark matter configurations of Fig. 1. To do this we add a time-varying Hernquist sphere \citep{Hernquist90} as an external baryon potential, $V_{\mathrm{bar}}$, at the center of the DM halos described by
\begin{equation} \label{eq:sn-evol}
V_{\mathrm{bar}}=- \frac{GM_{\mathrm{bar}}(t)}{r+b}, 
\end{equation}
with $b$ the scale radius and $M_{\mathrm{bar}}(t)$ the total baryonic mass at time $t$; the half-mass radius of the Hernquist potential is $r_{1/2}=b/(\sqrt{2}-1)$. We use $r_{1/2}=500$pc in most of our simulations being a typical size in bright Milky Way satellites. For this fixed $r_{1/2}$, we vary $M_{\mathrm{bar}}$ to span a range of magnitudes for the blowouts. We explore the  following SNe blowout feedback scenarios: 

\begin{itemize}
    \item[i)] Single blowout: starting from a soliton in equilibrium, 
    we add an external Hernquist potential of total mass $M_{\mathrm{bar}}$. 
    The mass of the external potential grows linearly with time until a maximum value $M_{\mathrm{bar}}$ is reached at $t=t_{\rm max}$, i.e. $M(t) = M_{\mathrm{bar}}t/t_{\rm max}$. The cycle starts at $t=0$ and the mass accretion phase reaches its maximum at $t_{\rm max}=200$ Myr, then remains constant for another 100 Myr. We then mimic a single blowout by forcing the external potential to instantaneously return to zero. At $t=300$ Myr the system evolves under no external potential. We follow the evolution of the system until 5Gyrs. The evolution of $M_{\rm{bar}}$ in the``bursty" phase is shown in the lower panel of Fig.~\ref{fig:fig1}. 

    \item[ii)] Multiple blowout: starting from a soliton in equilibrium, we repeat the same cycle as in the single blowout model for the subsequent blowouts. After each blowout, we evolve the system with no external potential for 200 Myrs before starting the next cycle, each cycle then has a period 500 Myr, mimicking several consecutive mass blowouts in the galaxy. We follow the evolution of 10 blowout cycles (5 Gyrs), after the blowouts we continue the simulations until 8 Gyrs. 

    \item[iii)] Soliton+halo: We initialize with the virialized SFDM halo in Fig. 1. We run two more simulations to study the case of one blowout and 10 multiple blowouts when a halo component in initially present. We compare our results with with the corresponding isolated soliton configurations.
\end{itemize}

As discussed in \cite{burger19}, to do a fair comparison between a single and multiple blowout scenario, we should compare the impact among the simulations that have the same total energy released from its  individual/multiple blowouts. 
For the external Herquist sphere, when the mass in each blowout cycle is set to zero the amount of energy 
$E = GM^2_{\rm bar}/(6b)$ is deposited into the halo. Since $E \propto M^2_{\rm bar}$ for a fixed $b$, the same energy will be released in $N$ less energetic episodes provided $M^{\mathrm{multi}}_{\mathrm{bar}}=M^{\mathrm{one}}_{\mathrm{bar}}/\sqrt{N}$, with $M^{\mathrm{one}}_{\mathrm{bar}}$ the mass blown out in one single episode. 

In our study we explored single blowouts of $M^{\mathrm{one}}_{\mathrm{bar}}=10^{6},10^{7}$ and $10^{8}\mathrm{M_\odot}$ and the corresponding multiple blowouts scenarios $M_{\mathrm{bar}}^{\rm multi}=M^{\mathrm{one}}_{\mathrm{bar}}/\sqrt{10}$. Hereafter, we will discuss the results for the two cases that show relevant impact on the dark matter, namely, the single blowout of $M^{\mathrm{one}}_{\mathrm{bar}}=10^8\mathrm{M_\odot}$ and 10 multiple blowouts of $M^{\mathrm{multi}}_{\mathrm{bar}}=\sqrt{10} \times 10^7\mathrm{M_\odot}$. We found that for the smaller mass blowouts  ($M^{\rm one}_{\rm bar} \leq 10^{7}\mathrm{M_\odot}$), the impact on the soliton is much smaller and the results are quite similar to the case in which blowouts are absent. We have also run two simulations for the latter baryon masses with $r_{1/2} $= 100 pc and found qualitatively similar results. For clarity, we will present the results for the runs with our fiducial value $r_{1/2} = 500$ pc, being more typical for the half-light radii of bright dwarf galaxies. 

\subsection{Numerical resolution and SP code implementation}

Given that our simple model considers blowouts which are ejected radially from the halo center, the symmetry of the problem allow us to simplify the study to one dimension. We therefore elected to implement it into an existing 1D Schr\"{o}dinger-Poisson solver in Mathematica: this code was previously tested and used in \cite{2022PhRvD.105j3506Z}, where it was found to match perturbative calculations and validated against a full 3D solver, {\sc chplUltra}. {\sc chplUltra} is a distributed pseudo-spectral Schr\"odinger-Poisson solver written in the Chapel coding language (\cite{ChapelChapterBalajiBook, ChapelIJHPCA}); details can be found in \cite{chplUltraPaper}. 

For all the configurations we discuss in this manuscript, we used our 1D code to
evolve the systems. A significant advantage to using 1D code is higher spatial
resolution: we use a box size of almost twice the virial radius $L = 76.7 \,
\rm{kpc}$ divided into $n = 2000$ bins for a spatial resolution of $\Delta r
\approx 0.038 \, \rm{kpc} \lesssim r_{vir}/1000$. 
We implement a hard outer boundary and have verified that for our choice of $L$, the boundary doesn't
appreciably affect the behavior of the central soliton core. 
We summarize the parameters for our initial configurations in Table \ref{tab:sims}.

\begin{table}
\caption{Parameters of the initial conditions for the scalar field dark matter simulations discussed in the manuscript. The columns are: (1) Scalar Field dark matter configuration (2)
Total mass of the configuration  (3) Soliton core radius; (4) Scale radius $r_{1/2}=b/(\sqrt{2}-1)$ of a Hernquist sphere that mimics the baryonic potential; (4) Mass ejected in a single blowout (5) Mass ejected in each of the ten blowouts; (6) Radial spatial resolution in our simulations.}
\setlength{\tabcolsep}{3pt}
\begin{tabular}{llccccc} 
 & Total Mass & $r_c$ & $r_{1/2}$ & $M^{\mathrm{one}}_{\rm bar}$  & $M^{\rm multi}_{\rm bar}$ & $\Delta r$
 	\vspace{1mm} \\ 
 & $[10^{8} \mathrm{M_\odot}]$ & [pc] & [pc] & $[10^{8} \mathrm{M_\odot}]$ & $[10^{7} \mathrm{M_\odot}]$ & [pc] \\
\hline
Isolated soliton & $2.97$ & 771 & 500 & 1  & $\sqrt{10}$ & 38pc\\
SFDM halo & $10$ & 771 & 500 & 1  & $\sqrt{10}$ & 38pc \\
\end{tabular}
\label{tab:sims}
\end{table}

\begin{figure*}
	\includegraphics[width=\columnwidth]{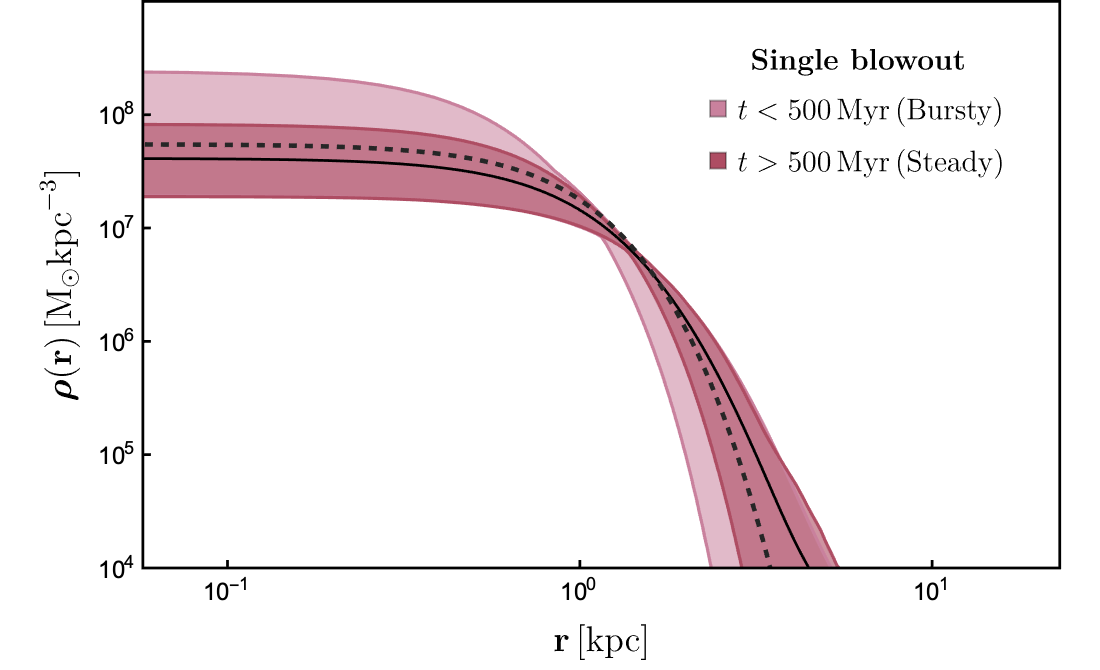}
    \includegraphics[width=\columnwidth]{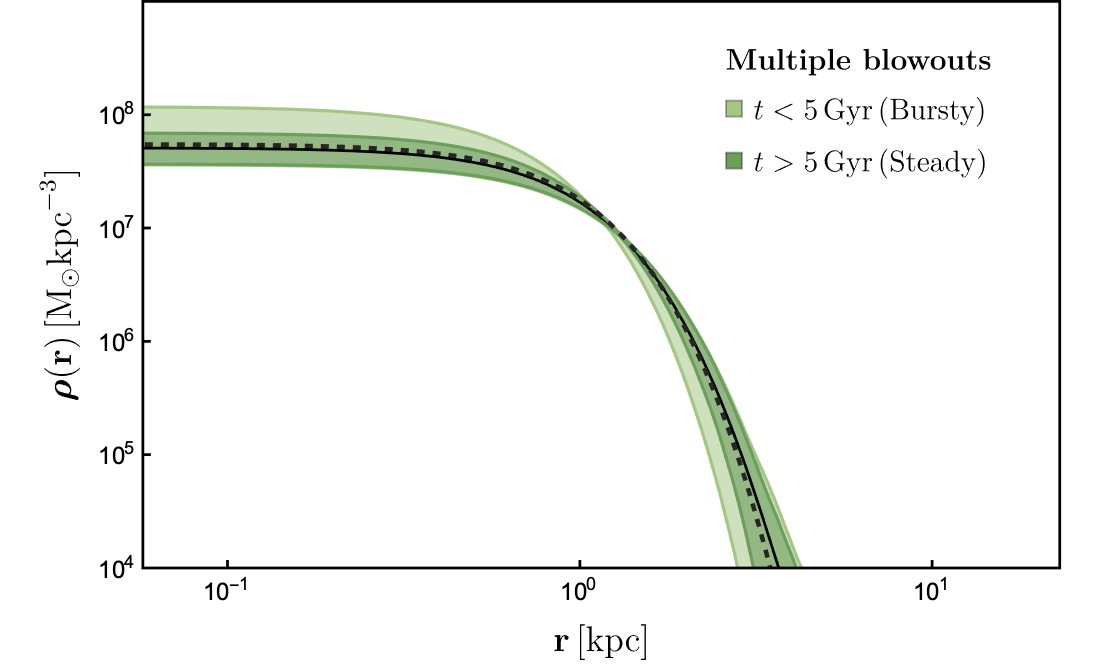}
    \includegraphics[width=\columnwidth]{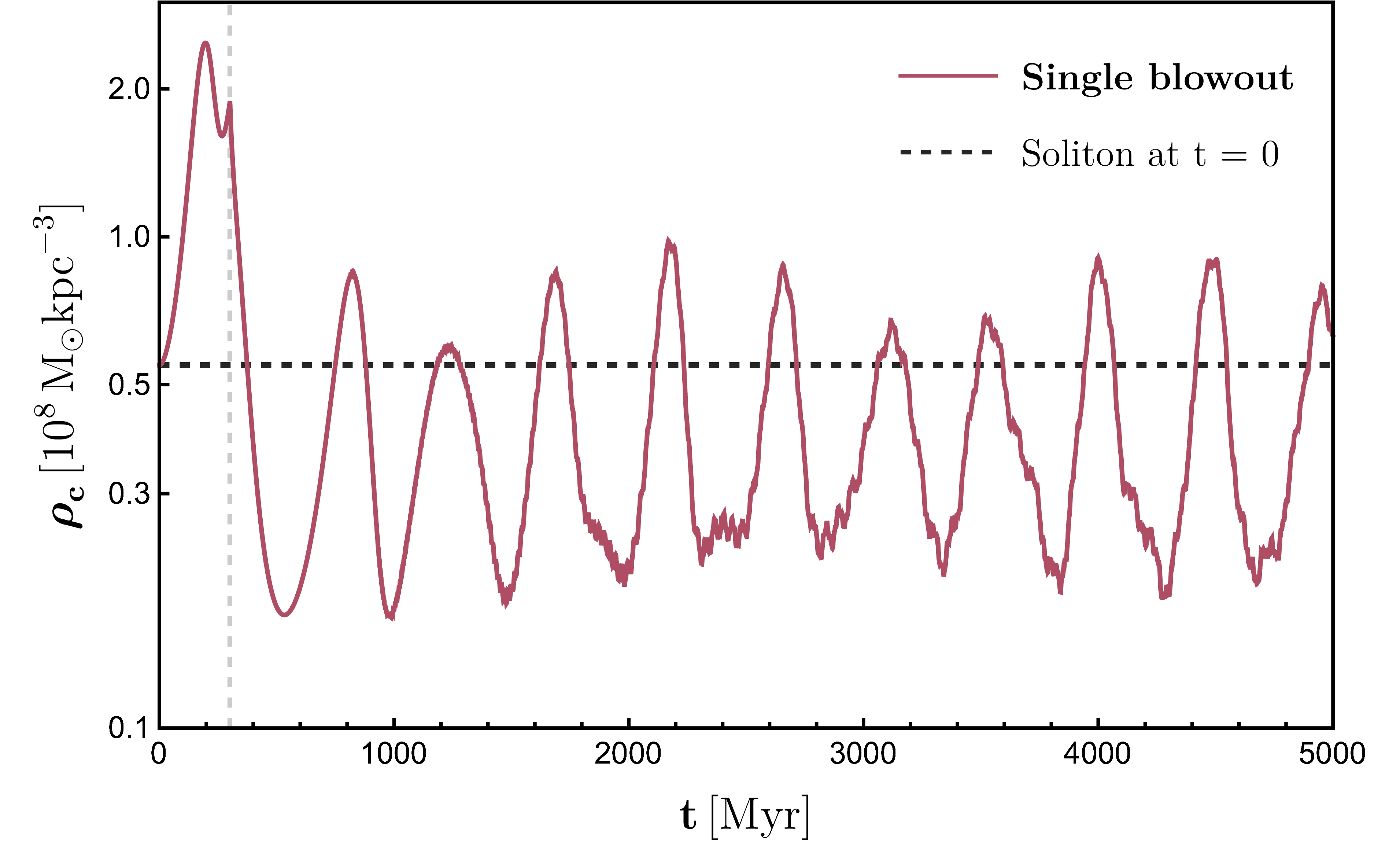}
        \includegraphics[width=\columnwidth]{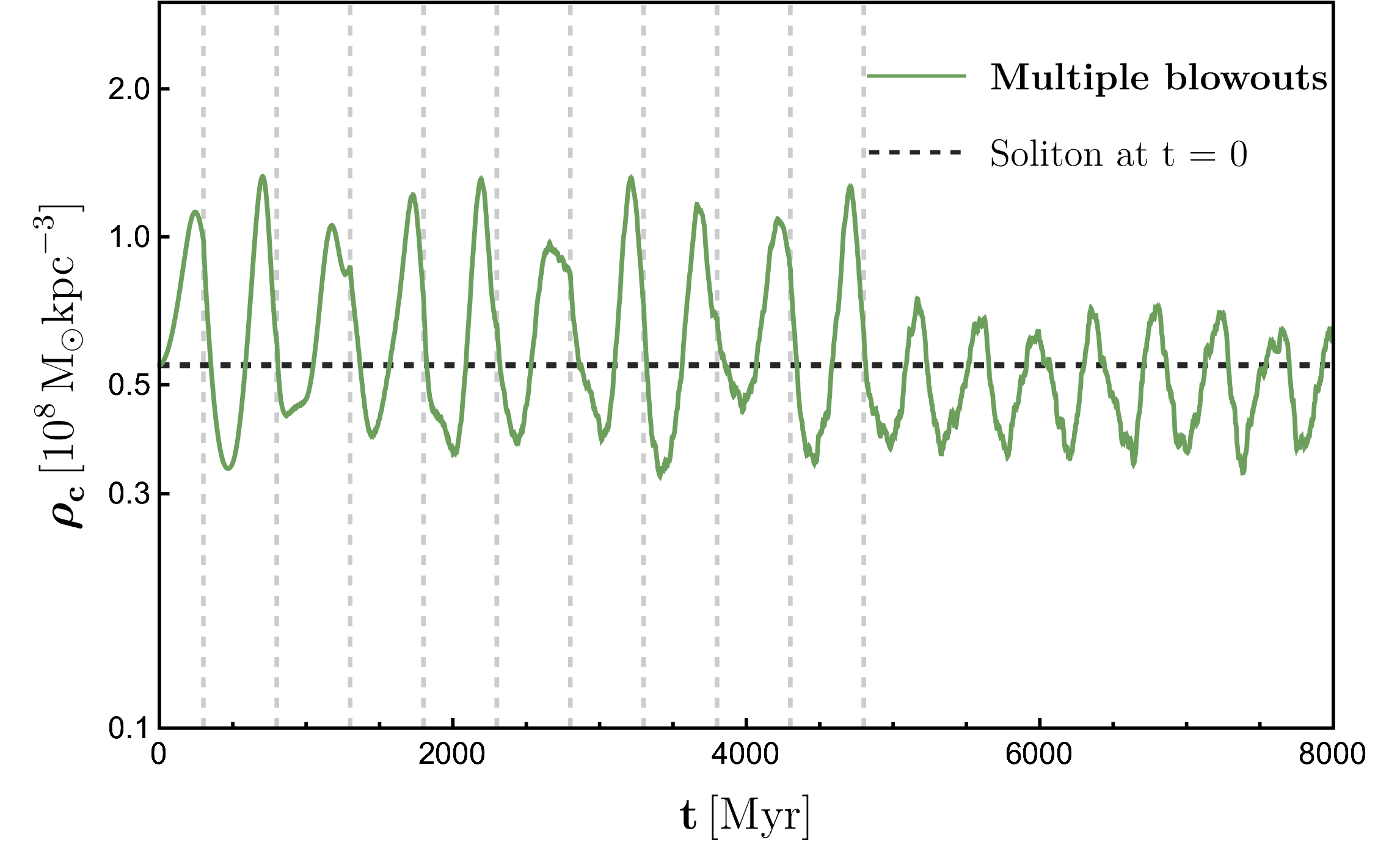}
            \includegraphics[width=\columnwidth]{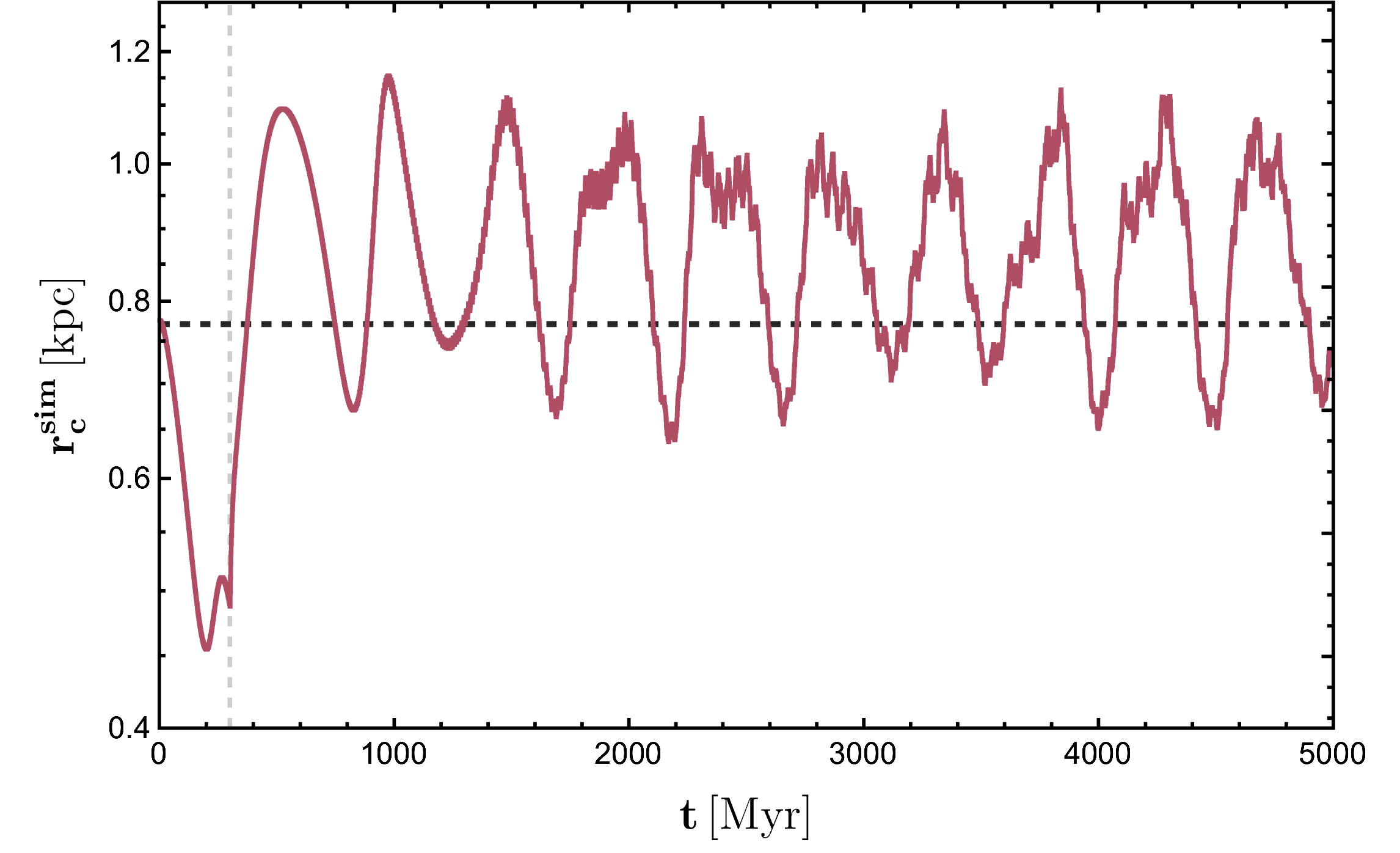}   
    \includegraphics[width=\columnwidth]{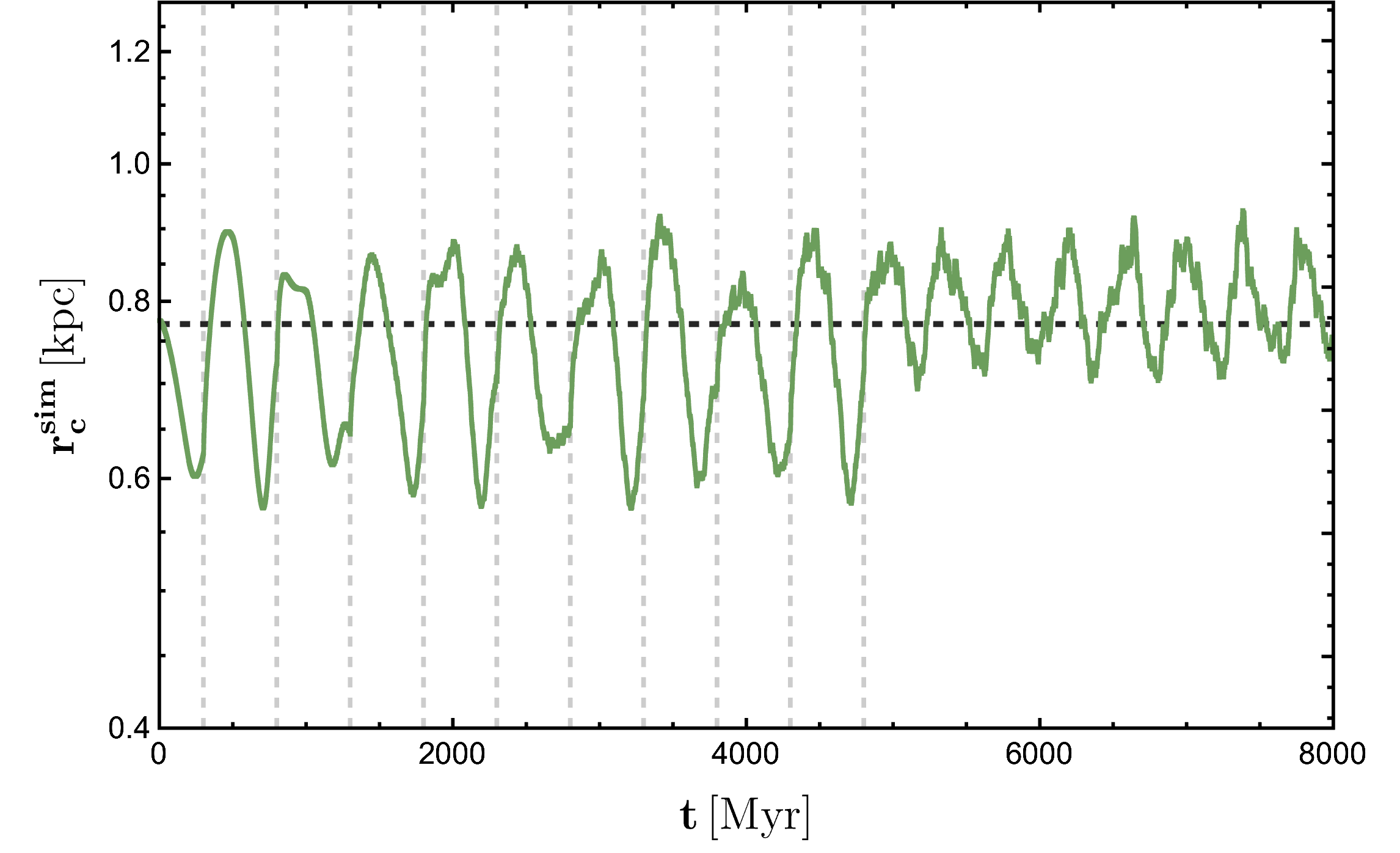}   
    
    \caption{We compare the evolution of an isolated soliton under single (left) and multiple (right) supernova SNe blowouts. The top row shows the radial dark matter distribution for the soliton profile in the bursty and steady phases (light and dark colors, respectively). The initial soliton profile is shown in the dashed line, and the average profile in the steady phase is shown in black. The middle row shows the time evolution of the solitons' central densities, with times of blowout indicated by the vertical lines and the initial value indicated by the horizontal line. The bottom row shows the evolution of the core radius, initially set to $r_c=0.771$ kpc.
    }
   \label{fig:fig2}
\end{figure*}

\begin{figure*}
    \flushleft 
    \includegraphics[width=\columnwidth]{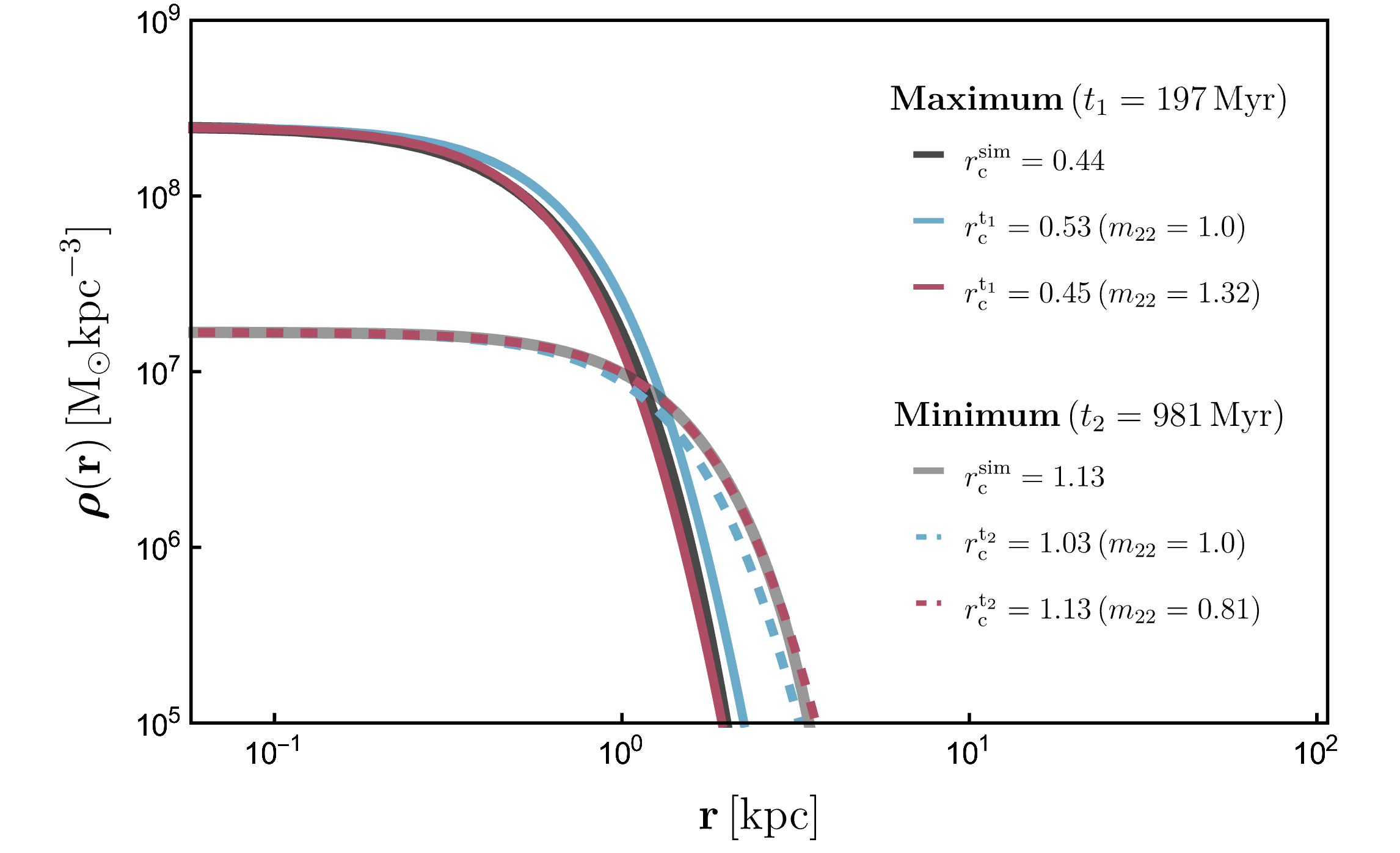}
    \includegraphics[width=\columnwidth]{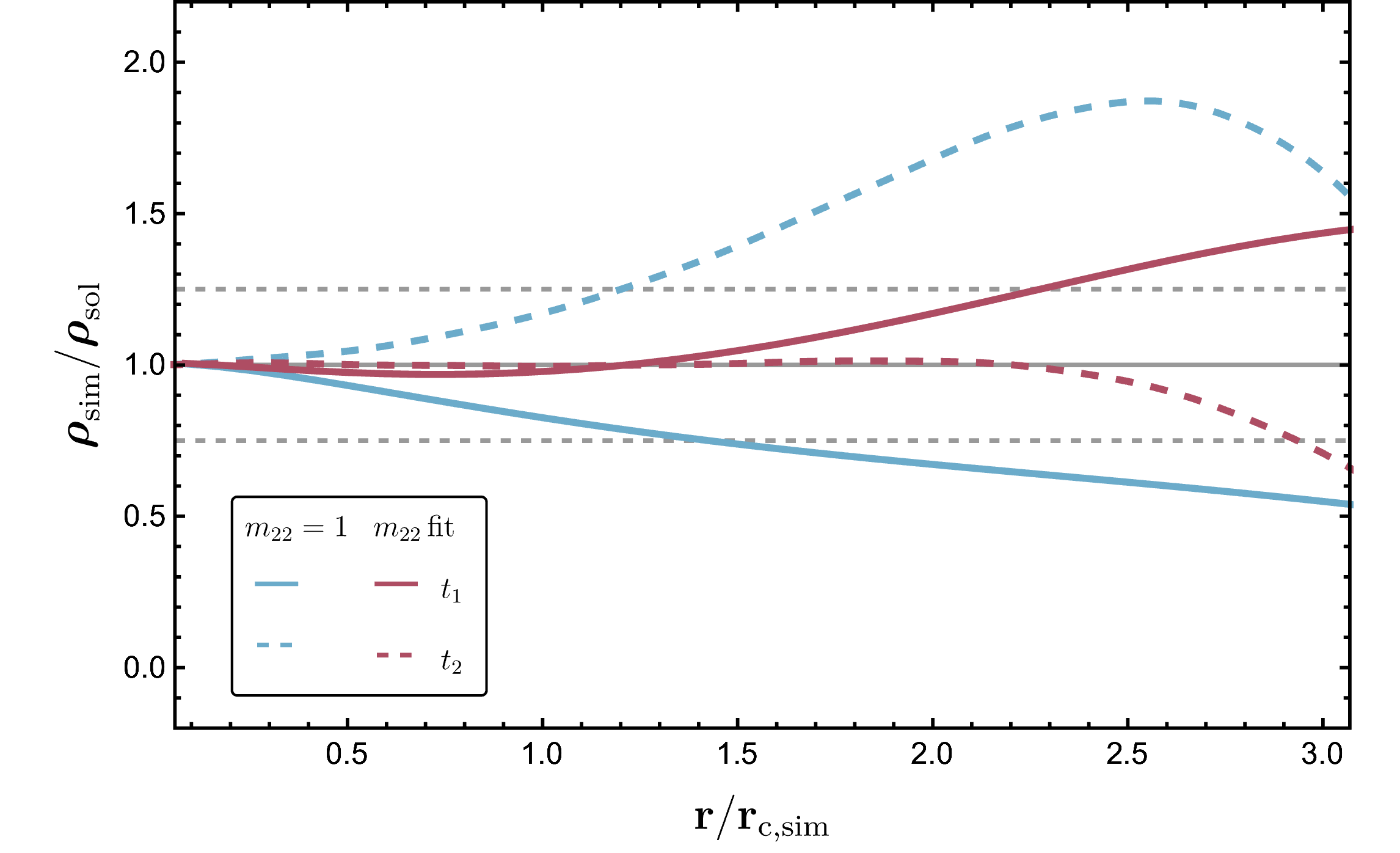} 
     
    \caption{Left: Dark matter density profile for the simulation (solid) with one $10^8 \mathrm{M_\odot}$ blowout at two different times t=197Myr (black) when the density is at a maximum, and t=981Myr (gray) when the density is at a minimum. We show for each time the comparison to the inferred soliton profile for $m_{22}=1$ (cyan) and the best-fit soliton (pink) obtained by assuming the central density ($\rho_c$) is equal to the value found in our simulation at the respective time. Right: Density ratio between the simulation and the two inferred soliton density profiles as a function of radial distance normalized to the simulation core radius at each time. We notice that there are times in the soliton evolution that it will be poorly described by a soliton in equilibrium with the correct boson mass, whereas it can still be well fitted by a soliton corresponding to a different boson mass. 
    }
    \label{fig:fig3}
\end{figure*}

\begin{figure*}
    \centering
        \includegraphics[width=\columnwidth]{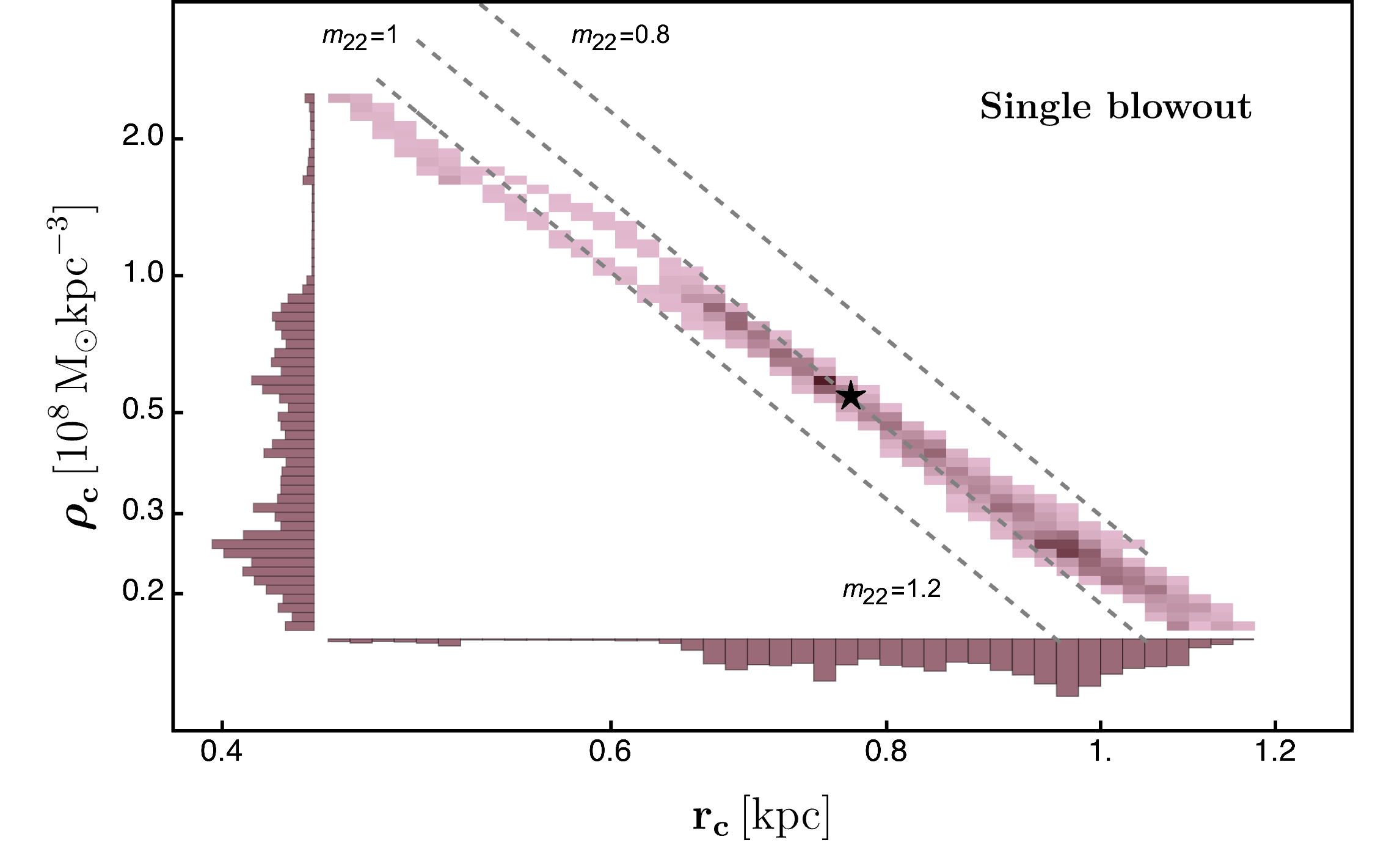}
     \includegraphics[width=\columnwidth]{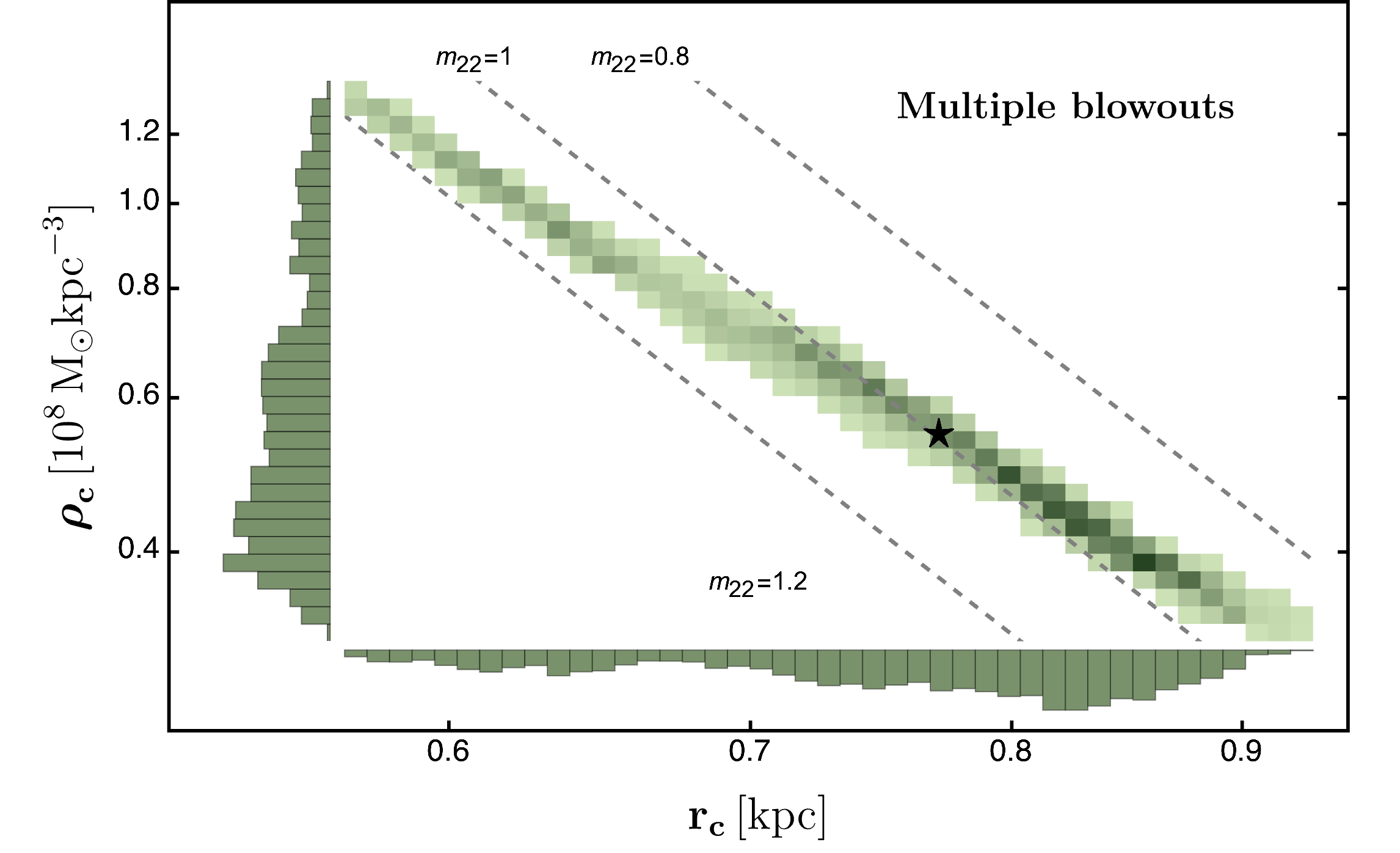}
        \includegraphics[width=\columnwidth]{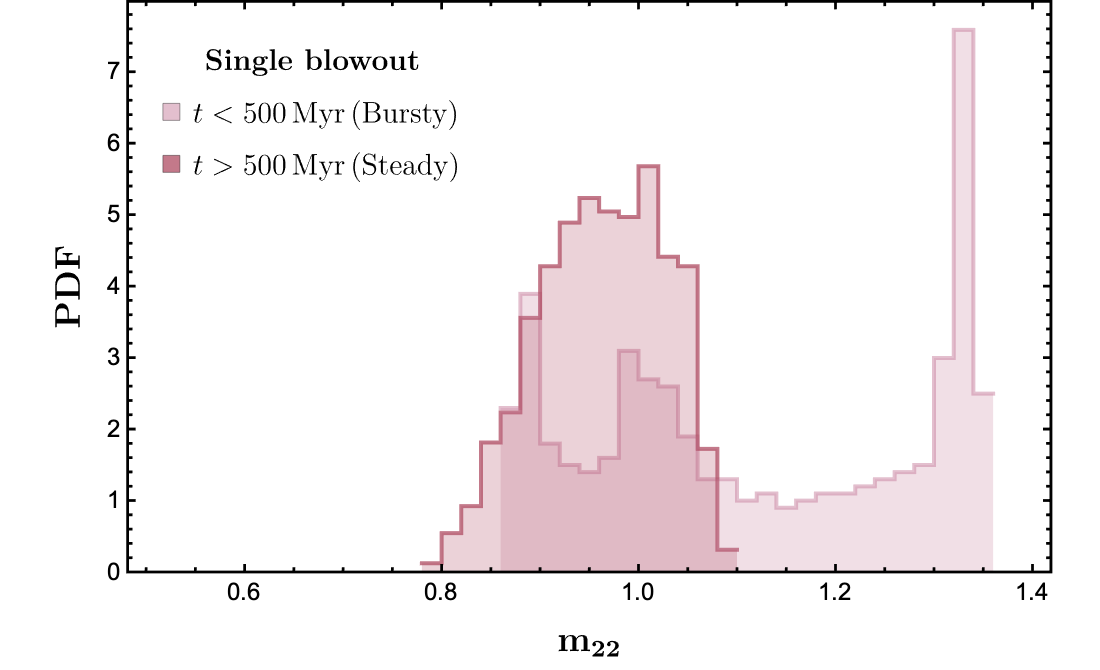}
         \includegraphics[width=\columnwidth]{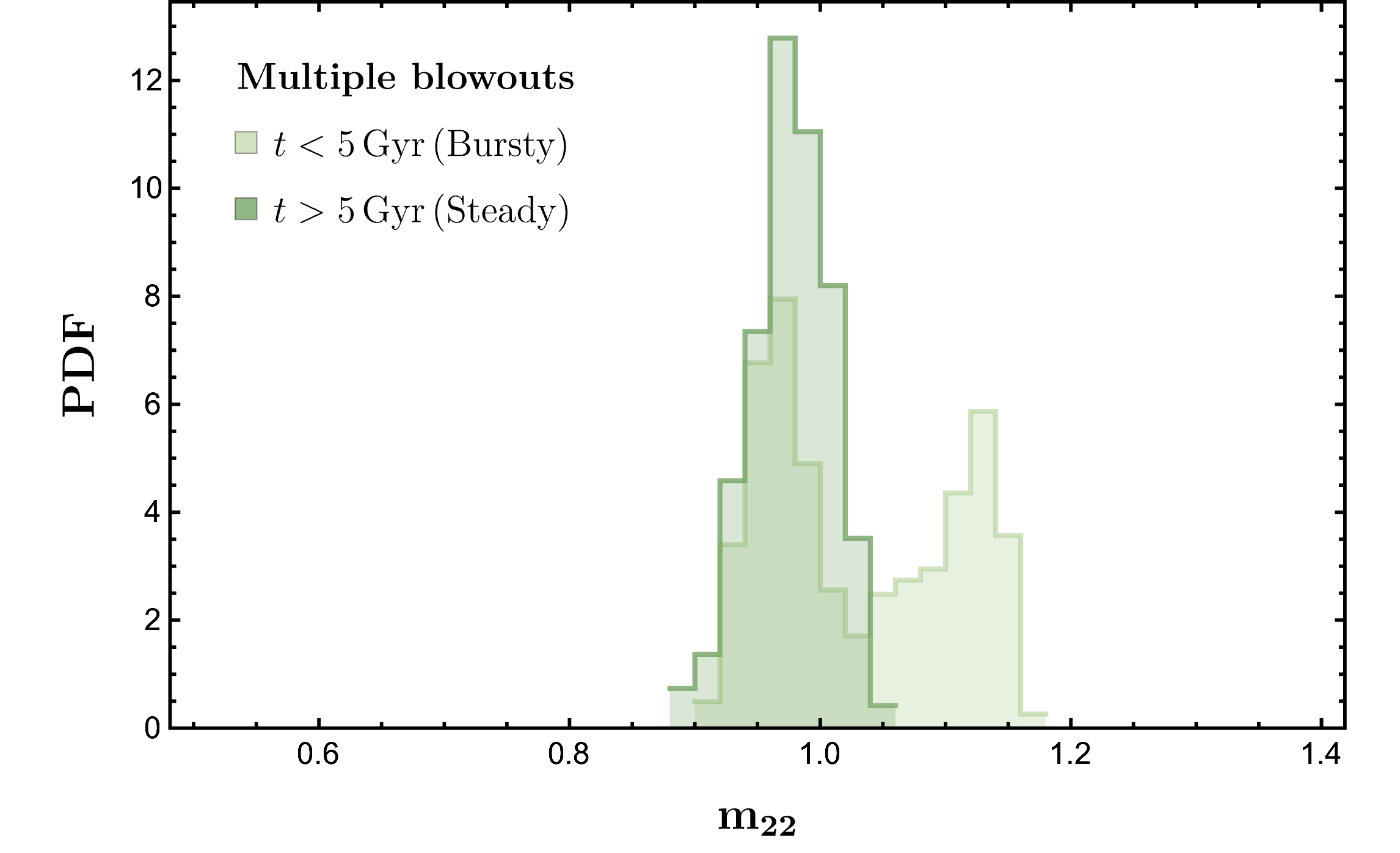}

    \caption{Top: Histogram of the evolution of the central density as a function of core radius for the simulation with one $10^8\mathrm{M_\odot}$ blowout (Left) and ten $\sqrt{10}\times 10^7\mathrm{M_\odot}$ blowouts (Right). Also shown (black star) are the initial values for the unperturbed soliton. Bottom panels show the PDF of the inferred scalar field mass for each time utilizing eq.(\ref{rhosol}) for each simulations as labeled.
    }
    \label{fig:fig4}
\end{figure*}

\begin{figure*}
    \centering
        \includegraphics[width=.33\textwidth]{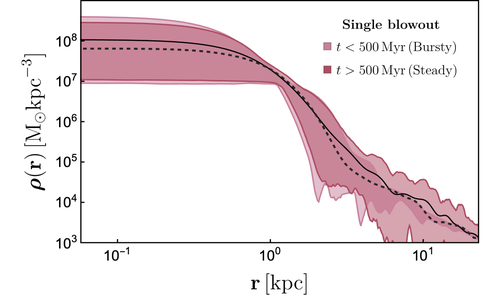}
          \includegraphics[width=.33\textwidth]{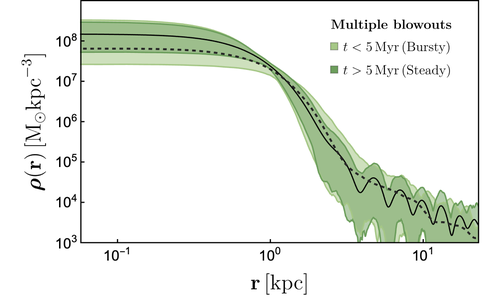}     
		\includegraphics[width=.33\textwidth]{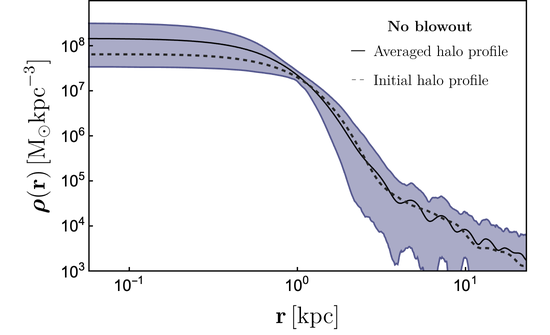}   

        \includegraphics[width=.331\textwidth]{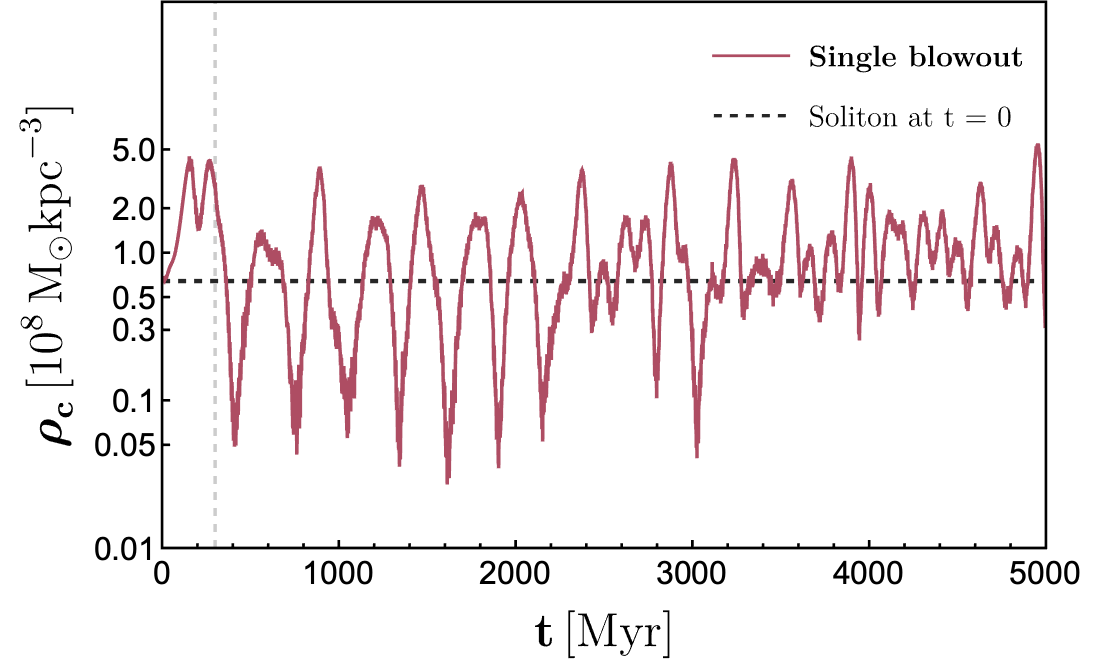} 
        \includegraphics[width=.331\textwidth]{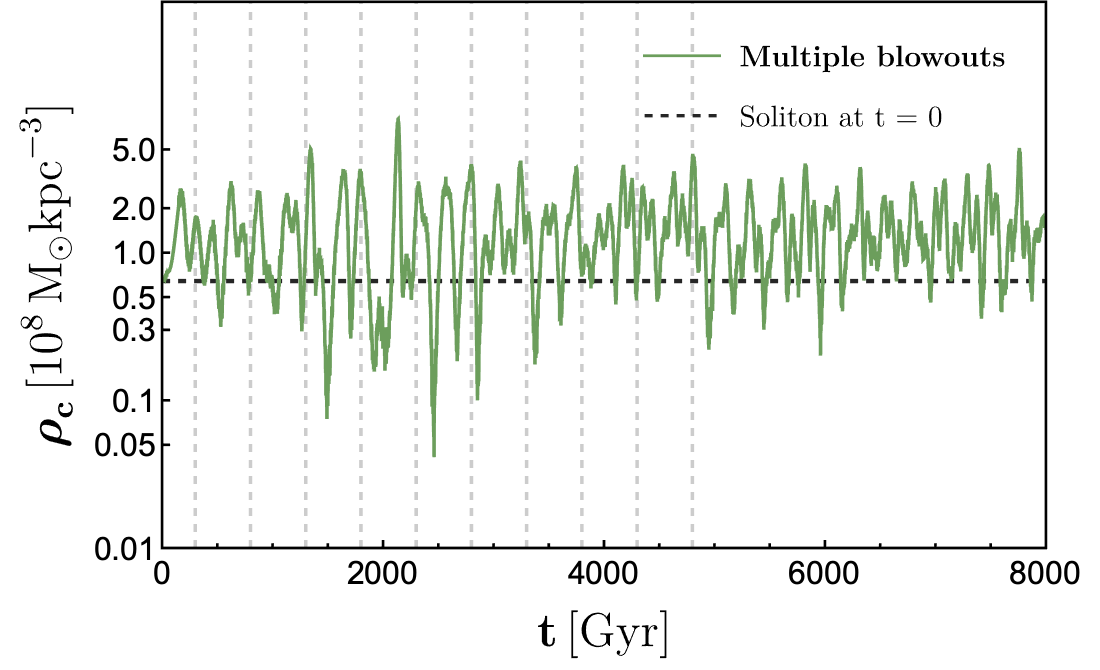}    
         \includegraphics[width=.33\textwidth]{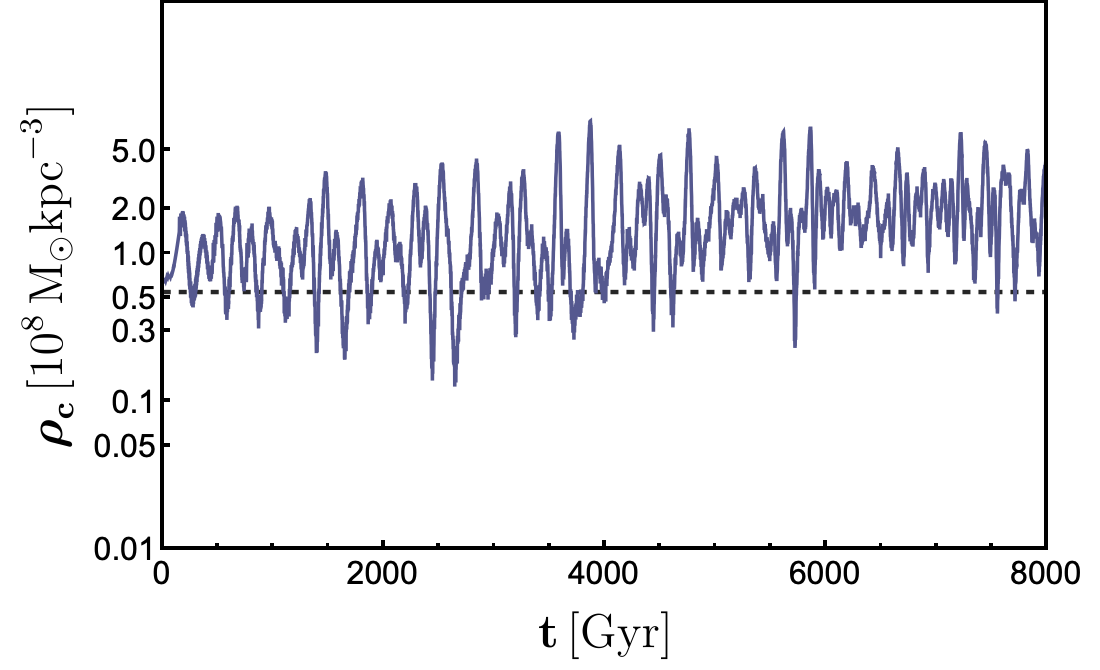}   
         
                 \includegraphics[width=.331\textwidth]{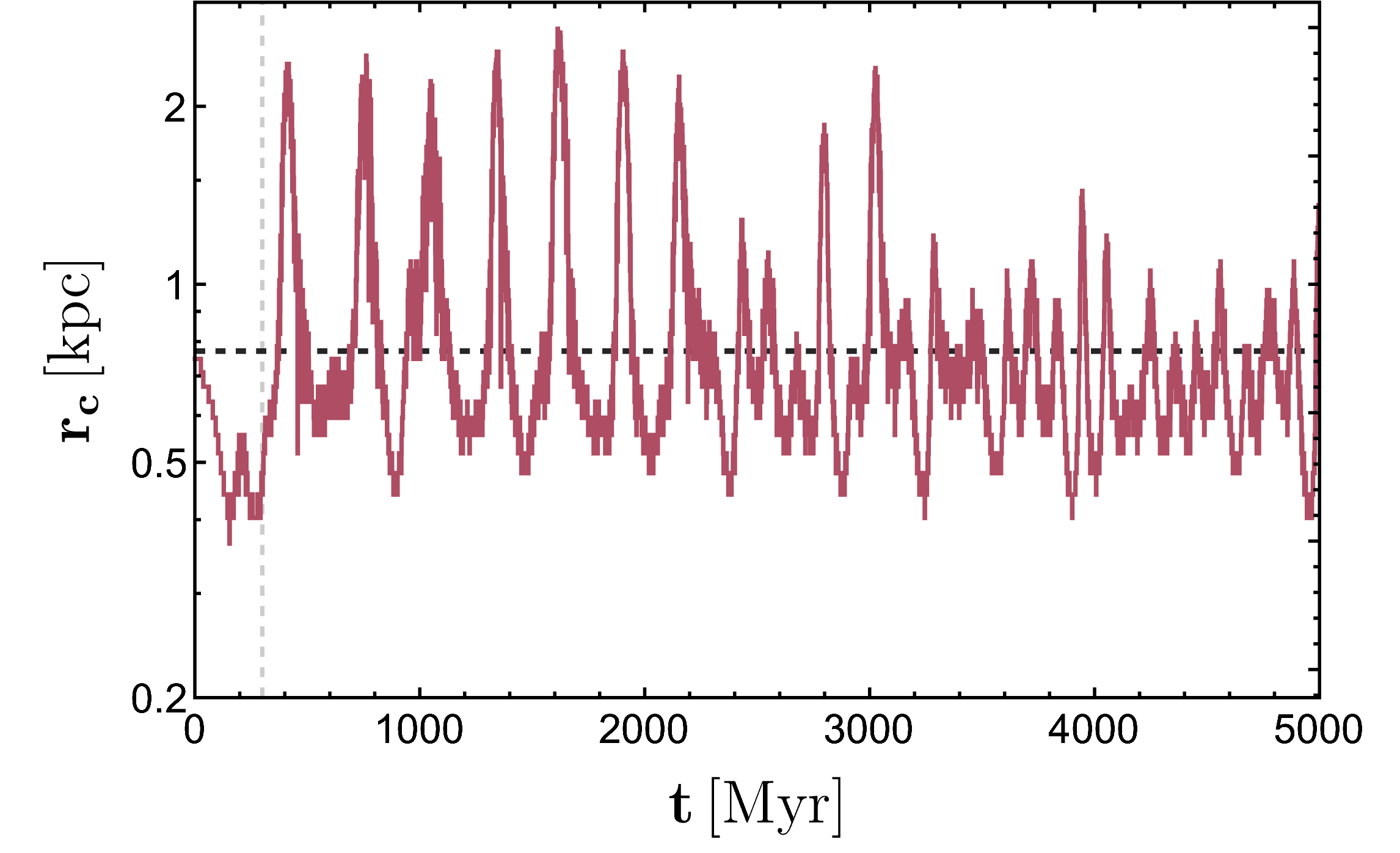} 
        \includegraphics[width=.331\textwidth]{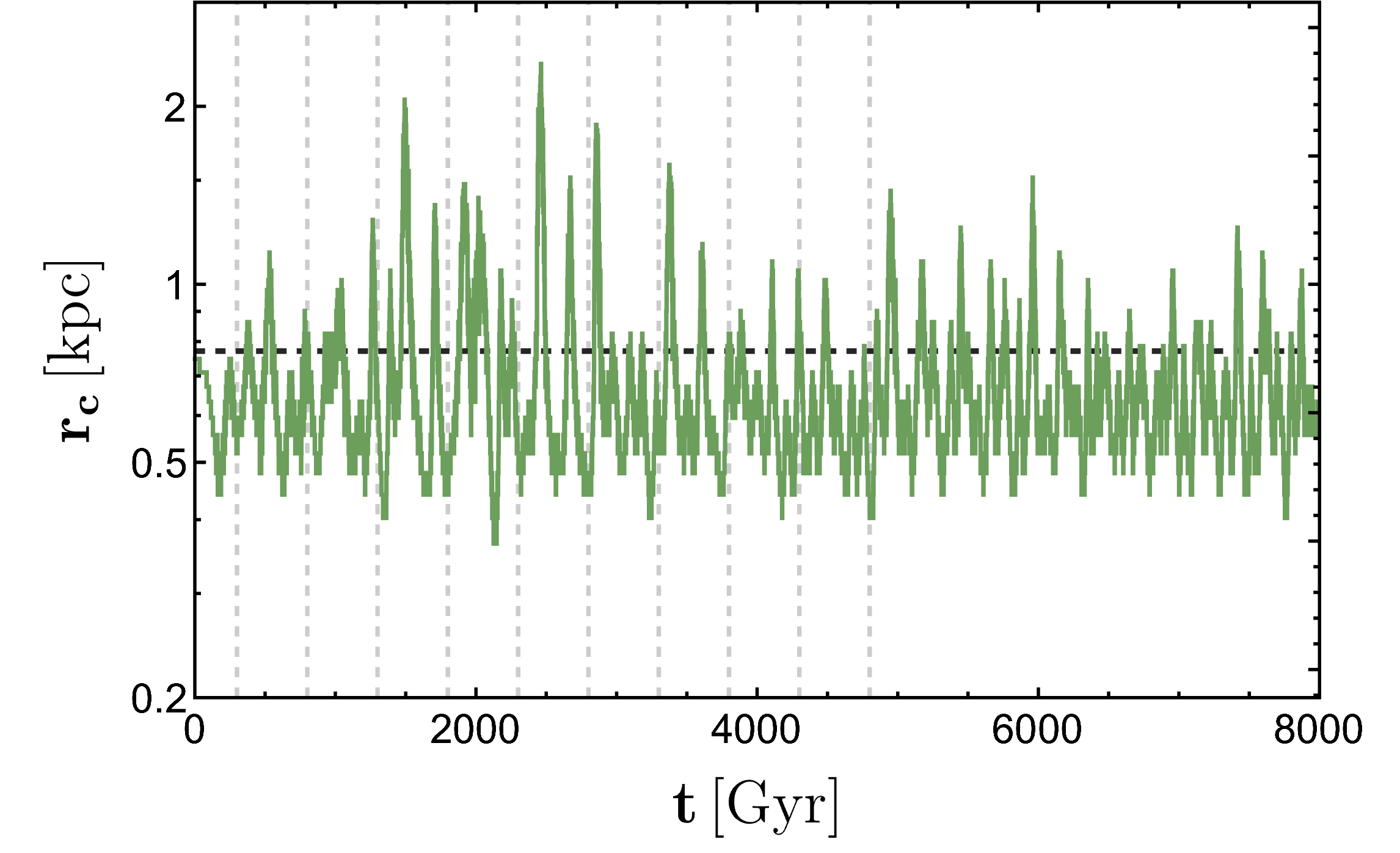}    
         \includegraphics[width=.33\textwidth]{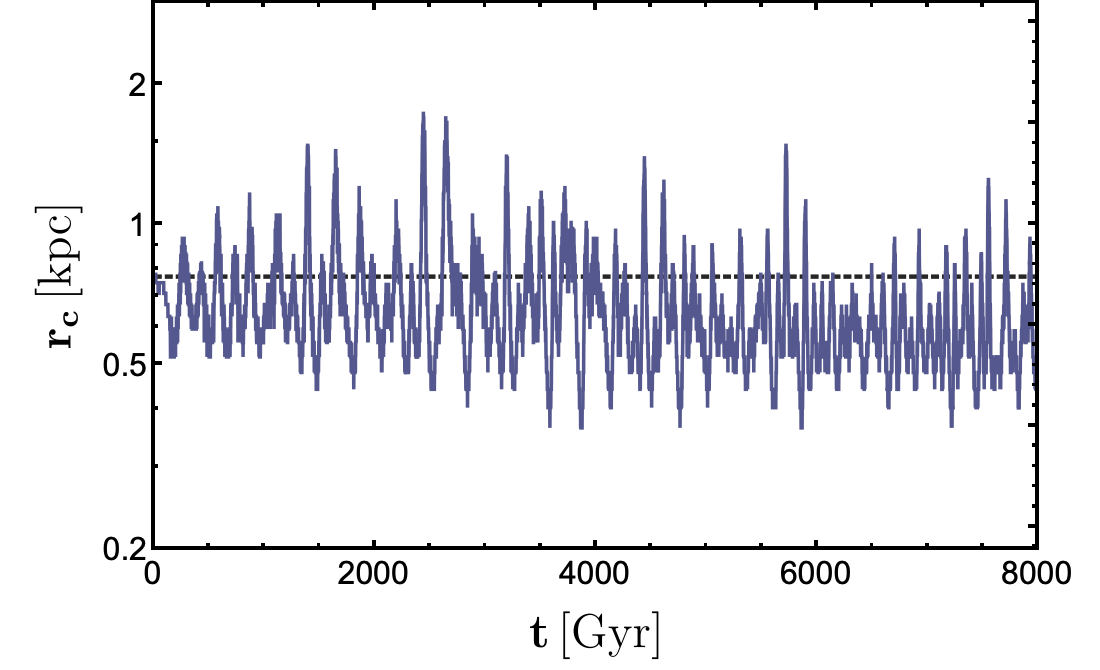}

       \includegraphics[width=.33\textwidth]{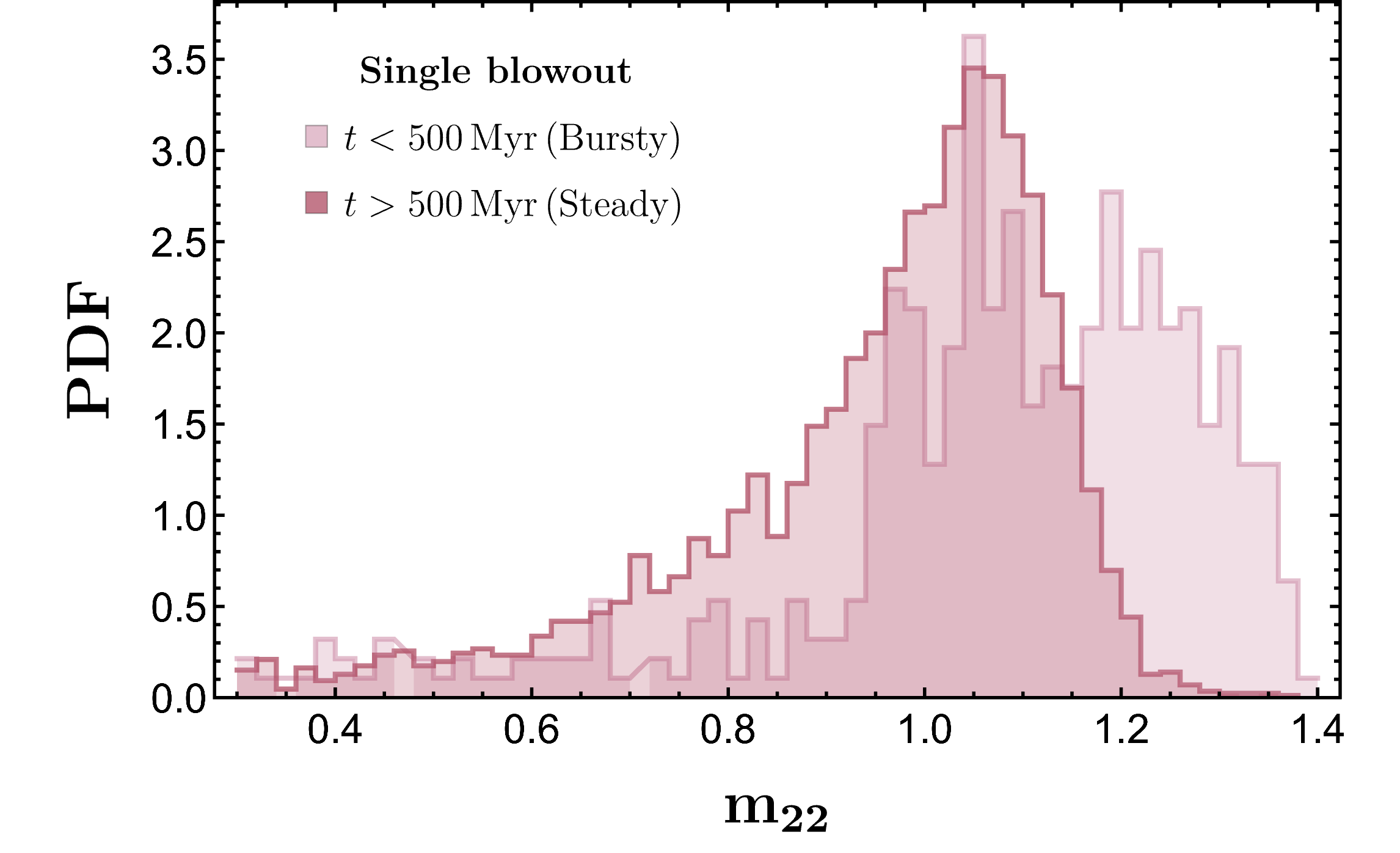}   
 \includegraphics[width=.33\textwidth]{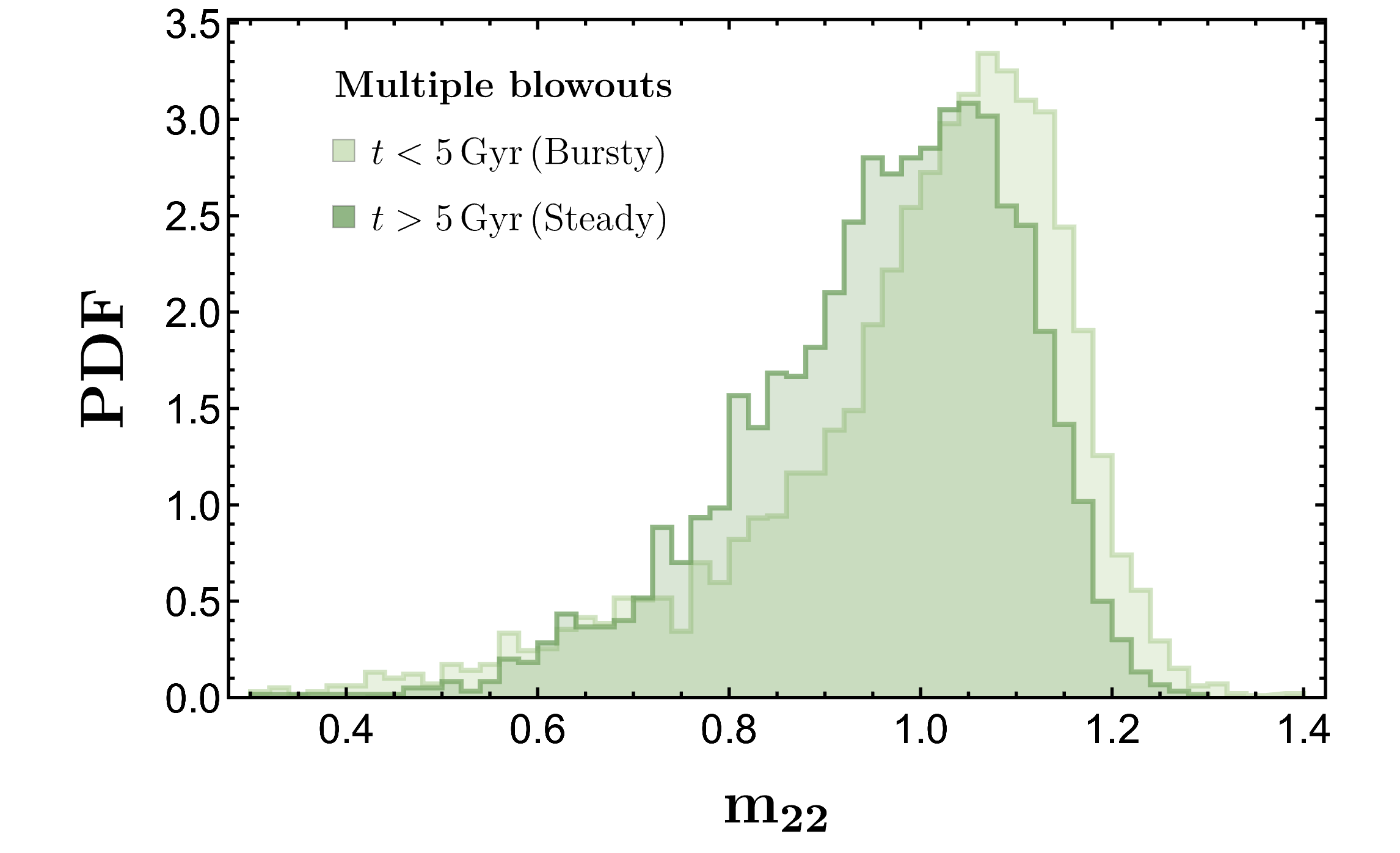}
   \includegraphics[width=.33\textwidth]{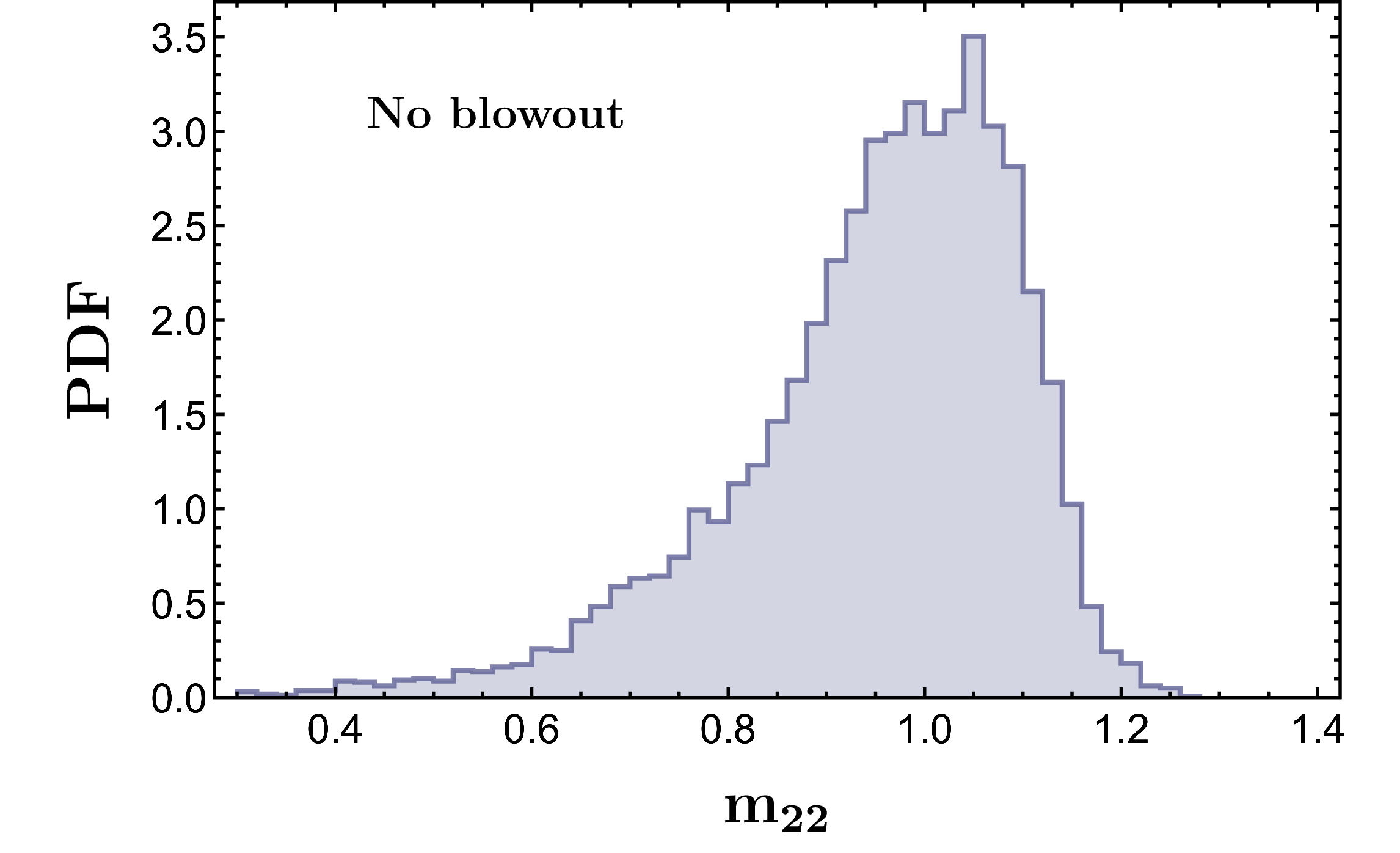}

     \caption{Shown (Top row) is the 5\%-95\% range of density profiles for the simulations that include an outer halo component and evolve under one blowout (Left column), ten blowouts (Middle column) and no blowouts (Right). Second row shows the time evolution of the central densities. Third row shows the evolution of the core radius and in the last row we show the PDF of the inferred SFDM mass for each simulation. 
     }
    \label{fig:fig5}
\end{figure*}

\begin{figure*}
	\includegraphics[width=\columnwidth]{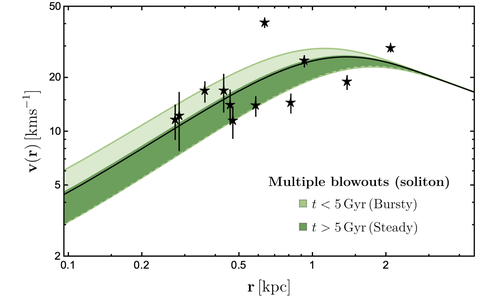}
	 \includegraphics[width=\columnwidth]{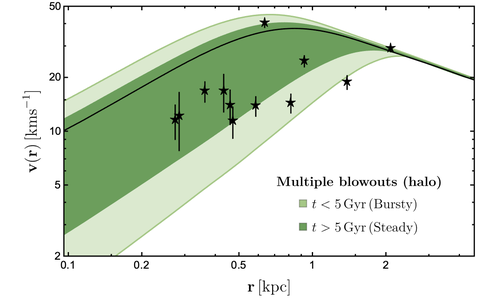}
    \includegraphics[width=\columnwidth]{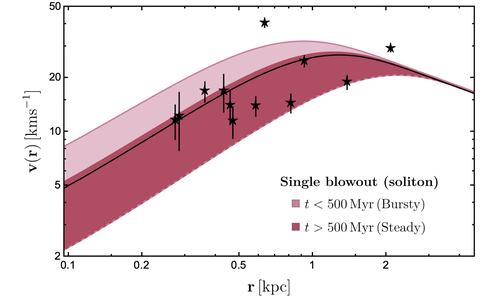}       
     \includegraphics[width=\columnwidth]{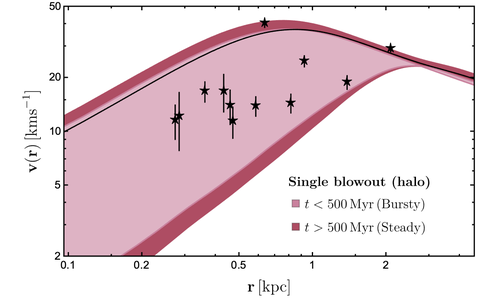}

    \caption{Range of circular velocity profiles for the simulations with an initial soliton (Left column) and soliton+halo (Right column). Top Row shows the evolution under ten blowouts while the case of one blowout is shown in the Bottom row. The shaded regions show the range between the median maximum and minimum values during the Bursty and Steady phases as described in the labels. Also shown are the estimated circular velocities at the half-light radii for Milky Way dwarf galaxies in the field (blue stars) obtained from applying the mass estimator in \citet{wolf10} and the galaxy data in \citet{garrison19} that was compiled from  \citet{mcconnachie12} (see references therein). The scatter in the galaxy data can be accounted for if dwarf galaxies are hosted by isolated solitons and had an early strong feedback episode followed by a steady evolution (lower left panel). If galaxies are hosted in solitons that retained their dark matter outer envelopes (right column), the observed galaxy scatter is consistent with the predicted range of our simulations but with much less dependence on the strength or the history of the blowouts (as shown in Fig. 5.)}
    \label{fig:fig6}
\end{figure*}

\section{Results}

\subsection{Evolution of isolated solitons under SNe blowouts}

For the isolated soliton configuration, Fig. 2 shows the change of the soliton density profile, soliton central density, and core radius for the single (multiple) blowout(s) with $M_{\mathrm{bar}}=10^8\mathrm{M_\odot} (\sqrt{10}\times 10^7\mathrm{M_\odot}$) during the blowout cycle (Bursty phase) and after the blowout cycle ends (Steady phase).

Comparing the effects of the SNe blowouts in these two cases, we find that in both simulations the external potential induces a large perturbation to the soliton central density (defined as the innnermost density in the simulation) and to the core radius, which is determined by the radius in which the dark matter density is half of its soliton central density $\rho(r_{c,\mathrm{sim}}) =\rho (0)/2$.  A single blowout ($M_{\mathrm{bar}}/M_{\mathrm{sol}} \approx 0.3$) leads to dark matter contraction during the time the external potential grows and settles; the blowout then triggers a large amplitude oscillation in the soliton density and core radius that remains until the end of our simulations with no signs of damping. The time dependence in the density distribution induced by the blowout inevitably implies a range of viable density profiles, even during the steady phase. 
In the top panel of Figure 2, we show the $1\sigma$ range of the density profile for the bursty and steady phases. We note that, as a consequence of the massive blowout, some of mass was removed from the center and the configuration size increased. This is reflected in the range of profiles during the steady phase that oscillate around a configuration of lower density and larger core radius than the initial soliton. 
For the case of multiple blowouts ($M_{\mathrm{bar}}/M_{\mathrm{sol}} \approx 0.1$), the large amplitudes in the oscillating central density remain as long as the blowouts are in effect. Notably, once in the steady phase the central core density and core radius substantially decrease their amplitudes, the resulting configuration after the blowouts closely oscillates around a new equilibrium configuration of a slightly lower density than the initial unperturbed soliton (top panel of Fig. 2). This small shift in density is likely due to some amount of mass ejected outside of the soliton resulting from the consecutive blowouts. 

These two configurations reveal that the magnitude of the SN feedback blowout can introduce large variation in the determination of the soliton density profile parameters at different times. In fact, since the SN blowouts induced large perturbations, it is not expected that the range of density profiles in Fig. 2 corresponds to different solitons in equilibrium that satisfy the scaling symmetry \eqref{lambdasym}. However, we have found that it is possible to describe the non-equilibrium configurations by solitons in equilibrium provided the boson mass $m_{22}$ is considered a free parameter. 

To illustrate this, in Figure 3 we compare the density profile in our simulations with the inferred solitons in equilibirum (Eq. \eqref{rhosol}) at two epochs for the single blowout case. We show the profiles at the time of the largest contraction (maximum density reached at $t$=197 Myrs) and at the first expansion (minimum density) after the blowout ($t$=981 Myrs).
As expected, the simulation profiles are poorly fitted by Eq. \eqref{rhosol} with a fixed boson mass $m_{22}=1$, we notice that even at $r \approx 1.2 r_{c,sim}$ the soliton densities already differ by $\geqslant 25\%$, the difference increases by up to a factor of $\sim 2$ at larger radius. If we now allow $m_{22}$ to be a free parameter and use Eq. \eqref{rhosol} to fit the simulation density profiles, we find much better fits that reproduce the simulation density profiles and differ by $<12\%$ within $r \approx 2 r_{c,sim}$, providing accurate estimations for the core radius. The latter fitted profiles can deviate by $<40\%$ at $r \approx 3  r_{c,sim}$, although differences at large radii are expected given that Eq. \eqref{rhosol} is limited to model a soliton in equilibrium for $r<3  r_{c,sim}$, near this radius the simulation profile could deviate from a soliton in equilibrium due to the interactions with the mass that is either being accreted or ejected by the continuously oscillating soliton.

In Fig. 4 we show a 2D histogram of the soliton core density and soliton core radius for the density profiles in our simulations, where we have included values at intervals of 1 Myr. For comparison we show the expected $\rho_c- r_c$ relation (Eq. \eqref{rhocore}) for solitons in equilibrium using our fiducial boson mass, and we also include the relations for different boson masses that enclose the range of values in the single and multiple blowout simulations.
We observe some scatter in the relations. As noted in Fig.3, due to the large amplitude of the density oscillation induced by the SNe feedback blowout, the parameters of the non-equilibrium simulated soliton can be described by equilibrium solitons that correspond to a different boson mass. In the single blowout the large amplitude leads to a broader histogram for $\rho_c$ and $r_c$, although we can still observe a larger concentration of points at $\rho_c \approx 0.28 \times 10^8\mathrm{M_\odot kpc^{-3}}$ and $r_c \approx 0.97$ kpc that correspond to the average soliton parameters in the steady phase. For the multiple blowout case, the smaller impact of SNe on the density amplitude results in a clear concentration of points around $\rho_c \approx 0.59 \times 10^8\mathrm{M_\odot} \mathrm{kpc^{-3}}$ and $r_c \approx 0.76$ kpc, similar to the initial unperturbed soliton. 

The time-dependent density profiles, induced in this case by the SNe blowouts, are a characteristic feature of the SFDM model. This feature has an important implication in the determination of the boson mass directly from the DM distribution in a given system. In the bottom panels of Fig. 4, we show the inferred boson mass for the single and multiple blowout cases derived from Eq. \eqref{rhocore}). Solving for $m_{22}$ given $\rho_c$ and the core radius $r_c$ from the simulations, we infer the mass for snapshots in intervals of 1 Myr and show the resulting PDF for the steady and the bursty regimes separately.
In both of these regimes, we observe the inferred values cluster near our true mass $m_{22}$=$1$ but have a non-zero width in their distributions. In the single blowout the boson mass range can be up to 30\% larger and 15\% lower than the true mass during the bursty phase, and 10\% larger and 25\% lower in the steady phase. For the case of multiple blowouts we find inferred boson masses that reach up to 15\% (10\%) larger and 10\% (5\%) lower values than our fiducial mass in the bursty (steady) phase.

The larger uncertainty in the inferred dark matter boson mass for a single blowout scenario can be attributed to our result that the dark matter density is more strongly perturbed by a single massive SNe blowout than 10 smaller blowouts, both scenarios deposit the same total energy into the dark matter. Therefore, trying to constrain the boson mass from dwarf galaxies that experience large starbursts or that have continuous starbursts episodes during their evolution will inevitably carry some degree of uncertainty that depends on the star formation history of each galaxy. For the smaller mass blowouts we simulated ($M^{\mathrm{one}}_{\mathrm{bar}}$=$10^6\mathrm{M_\odot}$) the uncertainty tends to be narrower and the boson mass estimation would become more accurate. 

\subsection{Impact of SNe blowouts in virialized SFDM halos}

In our present non-cosmological simulation, we define our SFDM halos as virialized gravitationally bound configurations that formed by relaxation of multiple soliton mergers. While the dark matter density of an isolated soliton decreases exponentially at large radii ($r>>3r_c$), the outer density will decrease less steeply in SFDM halos (see Fig. \eqref{fig:fig1}). Our initial SFDM halo has a soliton of the same mass as our isolated soliton simulations so that the main difference between these two initial configurations is in the interaction of the soliton with its outer halo.

In Figure \eqref{fig:fig5} we compare the response of our SFDM halo to the single $10^8\mathrm{M_\odot}$ and 10 multiple $\sqrt{10} \times 10^7\mathrm{M_\odot}$ blowouts, we have also included the evolution of the SFDM halo without blowouts. In this figure we show, from top to bottom, the $1 \sigma$ range and mean dark matter density profiles obtained by time averaging the time-dependent density profiles, the soliton core density, the soliton core radius evolution, and the inferred (via Eq. \eqref{rhocore})) boson mass distributions in each simulation, we separate the results in the steady and the bursty phases as indicated by the figure labels.

Contrary to the soliton case, even in the absence of blowouts there is some intrinsic oscillations in the soliton structure parameters that originate from the continuous interaction of the soliton with its outer dark matter envelope. As the simulations evolve, we observe some overall increase in the (largely variable) soliton central density of about a factor of $\sim 2$, a similar increase was noted for non-cosmological idealized systems in \cite{dutta21}. 
From the intrinsic scatter in the density profile with no blowouts, we find that there is a minimum scatter of $\approx$ 20\% around the mean boson mass inferred from the simulations (see bottom panel in Fig. 5). Notably, from the PDF of the boson masses we observe that in some cases the inferred mass underestimates the true mass ($m_{22}=1$) by $\sim 60\%$, although with low probability. Note that unlike a soliton, a halo would normally \textit{not} be entirely spherically symmetric, and would thus exhibit modes that are not captured in our 1D simulations. Thus, we would expect the scatter to persist in 3D simulations of halos, but the exact shape and width of the PDFs might change.

We observe that when the SFDM halo experiences the multiple consecutive blowouts there is a small increase in variability (larger scatter) for the density oscillation amplitudes during some of the bursty episodes, once it enters the steady phase, the overall evolution is indistinguishable from the case of no blowouts. The latter is also observed by their remarkably similar inferred boson mass PDFs. 

For the single blowout model, the SFDM halo has the strongest response to the blowout. After the blowout, similar to the soliton-only simulation, the central density shows large amplitude density oscillations for about $\approx 3$ Gyrs, it then fluctuates around a smaller average density than what is observed in the absence of feedback, implying the SNe blowout managed to eject some mass outside of the inner soliton. However, we note that as the configuration evolves further and the interaction between the inner soliton and its outer halo component slowly relaxes to a quasi-equilibrium state, the relaxation leads to damping of the fluctuation amplitudes. Towards the end of our simulation there is a small net increase in the core density, though it continues varying by a factor of  $\sim 2$, which are both common features in our (non-cosmological) SFDM halos. The different stages of the density evolution in our halo during the steady phase are reflected as  a large spread in the inferred boson mass PDF in Fig. \eqref{fig:fig5}, showing a slightly larger preference for smaller masses than the ten multiple blowouts run of equivalent injected energy or  the no blowout case. 

For our smaller SNe blowout simulation ($10^6\mathrm{M_\odot}$), we found behavior that is indistinguishable from the simulation when blowouts are absent (right column in Fig. \eqref{fig:fig5}). Recalling that for each blowout modeled by a Hernquist sphere the amount of energy that is deposited to the dark matter is $E_{\mathrm{blowout}}(M_{\mathrm{bar}})=GM_{\mathrm{bar}}^2/6b$, a factor of ten decrease in $M_{\mathrm{bar}}$ implies 100 times less energy deposited to the halo for a fixed value of $b$. Comparing the blowout energy to the potential energy of the soliton $W_{sol}$ between two of our simulations, e.g. a blowout of $M_{\mathrm{bar}}= 10^6_\odot$ and $M_{\mathrm{bar}} = 10^7\mathrm{M_\odot}$, we would have $|E_{\mathrm{blowout}}(10^6\mathrm{M_\odot})/W_{\mathrm{sol}}|\approx 10^{-2} |E_{\mathrm{blowout}}(10^7)/W_{\mathrm{sol}}|$, for the same initial soliton host. The rapid decrease in the strength of the blowout suggests that the resulting scatter in the inferred boson mass and soliton density profile would be similar below some blowout mass, here $\sim 10^6\mathrm{M_\odot}$; in this case the observed scatter would be mainly driven by the soliton-halo interaction rather than SNe driven.

We point out that in the simulations with a SFDM halo, the soliton is surrounded by an outer envelope which increases the central gravitational potential compared to the isolated soliton case. We estimated the difference by computing the ratio of the central gravitational potential of our two initial conditions and find that it amounts to $\approx20\%$---that is, $|E_{\mathrm{blowout}}(M_{\mathrm{bar}})/W_{\mathrm{sol}}|\approx |E_{\mathrm{blowout}}(M_{\mathrm{bar}})/W_{\mathrm{halo}}|$. Therefore, disrupting the center of the soliton in the presence of a SFDM halo requires only a mild increase in the blowout mass compared to the case where the soliton is in isolation.

\subsection{Comparison with circular velocities of field dwarf galaxies in the Local Group }

In the previous sections we found that for SFDM configurations, SN-driven blowouts induce a rapid change in the parameters that determine the central soliton density; this changing density results in a range of viable soliton configurations that describe the evolution of the system. In Figure \eqref{fig:fig6}, we show the  range of circular velocity profiles for the isolated soliton (left column) and the virialized SFDM halo (right column) derived from the associated range of density profiles in the simulations with a single $10^{8} \mathrm{M_\odot}$ (top row) blowout and multiple $\sqrt{10} \times 10^{7} \mathrm{M_\odot}$ blowouts (bottom row). 
The SFDM configurations have maximum circular velocities in the scale of $V_{\mathrm{max}} \approx 30$ $\mathrm{km s^{-1}}$ (the associated soliton mass $\approx 3 \times 10^{8}\mathrm{M_\odot}$ for $m_{22}$=1), which is the relevant scale for the small-scale CDM issues. 
We compare with field dwarf galaxies in the Local Group. The star-shaped data points in Fig. \eqref{fig:fig6} show the inferred circular velocities at the respective galaxy half-light radius given by $V_{1/2}\equiv (GM_{1/2}/r_{1/2})^{1/2}$, where $M_{1/2}$ is the inferred dynamical mass at the galaxy half-light radius, $r_{1/2}$, we use the values compiled and reported in \cite{garrison19} \footnote{The observational data are from  \cite{mcconnachie12,kirby14,simon07,fraternali09,collins13,Leaman_2012}.}(see references therein).

From the observed range of circular velocity profiles in the SFDM configurations in Fig. \eqref{fig:fig6} we find that a single massive blowout (of a third of the soliton mass) seems to be more effective in driving dark matter out of the soliton than multiple repetitive blowouts (here ten cycles releasing 10\% of the soliton mass each) of equal total energy released. Having more mass ejected from the center results in a general increase in the rotation curve range in the single episode cases compared to the multiple-episode scenarios.

In Fig. \eqref{fig:fig6}, we compare the circular velocity profiles obtained from our set of simulations to the current observational constraints on $M_{1/2}$ for isolated bright dwarf galaxies in our Local Group. 
For the isolated soliton simulations, we find that our simulated isolated soliton of $V_{\mathrm{max}} \approx 30 \mathrm{km s^{-1}}$ under one single massive blowout is consistent with most of the data, provided we account for the significant scatter in the rotation curve driven by the single strong blowout. If the same total energy is now injected to the dark matter over 10 multiple episode, the predicted range of circular velocities is reduced, implying less intense feedback episodes yield solitons that are more resilient to feedback. Even in this case, we still see a good agreement with half of the dwarf galaxy data points. We conclude from these results that if galaxies are at present hosted by soliton-dominated halos, galaxies that had strong starbursts in the past (which might be reflected in their star formation histories) would imply a larger uncertainty in their inferred dark matter mass distributions. 

In the case of our SFDM halo simulations, we find remarkably that  both models of SNe feedback blowouts result in a predicted range of circular velocities that are in great agreement with the observed galaxy data. We additionally note that both feedback scenarios, the single and multiple SNe feedback blowouts, yield significant scatter in the dark matter densities and inferred circular velocities. In fact, the scatter in the density seen in the cyclic feedback blowout model is comparable to that of the SFDM simulation in the absence of feedback. Therefore, our simulations seem to indicate that the large uncertainty in their dark matter circular velocity profiles is mainly driven by the strong soliton-halo interaction, being only mildly dependent on the SNe blowout mass (at least within our explored blowout range, $M_{\mathrm{bar}}<0.3M_{\mathrm{sol}}$) or the total number of SNe starbursts events that occur in the galaxy lifetime.

It is interesting that the intrinsic variability of the SFDM dark matter central distribution can account for the wide diversity of of dynamical mass measurements at the half-light radii in Local Group field dwarf galaxies. Our simulations reveal that the impact of violent SNe effects on SFDM configurations is not trivial and can largely modify the structure of the inner soliton at different times throughout its evolution, even if in the absence of strong feedback activity. From the results in this work, we conclude that in order to improve the accuracy of the constraints to the  fundamental SFDM model parameter (i.e. the boson mass) when baryonic feedback effects are included, it is important that we use observations that go beyond the properties of a galaxy at a single time in its evolution. Having information of the galaxy's past history can help to identify the range of viable dark matter distributions and in turn put stronger constraints to the inferred boson mass. To fully account for the baryonic and dark matter interplay we would require full hydrodynamics simulations in the SFDM paradigm,  we leave such study for the future.

\section{Conclusions}

In this manuscript we have performed high-resolution numerical simulations in the context of Scalar Field Dark Matter with a boson mass $m=10^{-22} \rm \mathrm{eV/c^2}$ to study the impact of supernovae feedback in the form of instantaneous mass blowouts on a $V_{\mathrm{max}} \approx 30 \mathrm{km  s^{-1}}$ scalar field dark matter self-gravitating configuration hosting a typical isolated bright dwarf galaxy.
We simulate two types of scalar field dark matter distributions: an isolated single soliton and a SFDM halo, the latter described by a soliton embedded in an extended dark matter envelope. We explore the impact of SNe feedback blowouts by including a time-dependent external potential at the center of the SFDM distribution. To explore how the dark matter density at the center responded to the various feedback strengths, we varied the magnitude of the blowout mass and the number of blowout episodes.

From our simulations the general conclusion is that a single blowout event leads to a stronger perturbation to the soliton density than 10 multiple blowouts imparting equal total energy to the halo. The same qualitative conclusion holds whether the configuration is an isolated soliton or a SFDM halo.
Interestingly, for a CDM halo it is also the case that a single episode has the strongest effect on lowering the central density; however, erasing the dense cusp at the center of a classical Milky Way dwarf galaxy typically requires massive ($\geqslant 10^8\mathrm{M_\odot}$) blowouts, while smaller blowouts ($\sim 10^7\mathrm{M_\odot}$) still result in a cusp \citep{shea13,burger19}. In the present work, we focused on studying how the central soliton of a dwarf SFDM halo would respond to SNe blowout feedback events in a similar regime of blowout masses.

More specifically, we found that for an isolated soliton of mass $2.96 \times 10^8\mathrm{M_\odot}$, blowing out a total mass of $10^8\mathrm{M_\odot}$ in one single event will induce a long-lasting oscillatory perturbation in the soliton core density with values that differ by a factor of 2. For the same amount of energy imparted to the system but distributed in 10 multiple blowouts,  each ejecting $\sqrt{10} \times 10^7\mathrm{M_\odot}$ onto the dark matter, the central density amplitude fluctuates within 20\% of the mean value. For the SFDM halo simulations, we find that the central soliton experiences a continuous interaction with its outer dark matter envelope that yield a constantly fluctuating central density, this non-vanishing perturbation leads to a system in pseudo-equilibrium with central soliton densities that can rapidly vary by factor of 2 or larger. 
In these halo configurations, the net impact of the SNe blowouts is to enhance the range in which the central density fluctuates compared to the simulation with no feedback. The range is only mildly affected under ten repetitive $\sqrt{10} \times 10^7\mathrm{M_\odot}$ blowouts, but it can reach lower densities under the impact of a single massive blowout of $10^8\mathrm{M_\odot}$. 

We find that the perturbed central dark matter distribution from the simulations can still be well-fitted by the analytic soliton profile from Eq. \eqref{rhosol} with good accuracy within the core radius; this is true provided we consider the boson mass as a free parameter in fitting equation, in addition to the core radius. These density oscillations are a characteristic feature of SFDM configurations, we found that an important consequence of the constantly evolving structures is that attempting to infer the boson mass from fitting the state of the dark matter structure at a single time would generally yield a value that is at most $\sim 20\% $ accurate to the true boson mass. Configurations that undergo stronger feedback effects result in larger uncertainties. 

Comparing the inferred circular velocities from our SFDM halo simulations with the velocity measurements at the half-light radius of bright dwarf galaxies in the Local Group, we find that currently dwarfs are consistent with being hosted by a $V_{\mathrm{max}} \approx 30 \mathrm{km s^{-1}}$ dwarf scalar field dark matter halo whose soliton has a mass (for $m_{22}=1$) of $3\times 10^{8}\mathrm{M_\odot}$ and has an induced time-dependent density profile originating from the quantum nature of the dark matter and its response to baryonic feedback. 
The time-dependence in the soliton core results in an scatter that is able to accommodate the diversity of observed dwarf sizes.

In order to understand whether the diversity in the stellar properties of dwarf galaxies can be explained under the SFDM paradigm, we require self-consistent SFDM cosmological hydrodynamics dwarf galaxy simulations that account for not only the gravitational impact of SNe feedback but the complex interplay between the dark matter and the various astrophysical effects that play a role in determining the structure of a galaxy. We leave this endeavor for a future work.

\section*{Acknowledgments}

While at Yale, JZ was supported by the Future Investigations in NASA Earth and Space Science and Technologies (FINESST) grant (award number 80NSSC20K1538). Research at Perimeter Institute is supported in part by the Government of Canada through the Department of Innovation, Science and Economic Development Canada and by the Province of Ontario through the Ministry of Colleges and Universities.
V.H.R. acknowledges support by the YCAA Postdoctoral Prize Fellowship.

\bibliography{mybib}

\begin{thebibliography}{73}
\expandafter\ifx\csname natexlab\endcsname\relax\def\natexlab#1{#1}\fi

\bibitem[{Bar {et~al}\mbox{.}(2018)Bar, Blas, Blum, \& Sibiryakov}]{bar18}
Bar N., Blas D., Blum K., Sibiryakov S., 2018, Phys. Rev. D, 98, 083027

\bibitem[{Benítez-Llambay {et~al}\mbox{.}(2019)Benítez-Llambay, Frenk,
  Ludlow, \& Navarro}]{Benitez19}
Benítez-Llambay A., Frenk C.~S., Ludlow A.~D., Navarro J.~F., 2019, Monthly
  Notices of the Royal Astronomical Society, 488, 2387

\bibitem[{Bose {et~al}\mbox{.}(2019)Bose, Frenk, Jenkins, Fattahi, Gómez,
  Grand, Marinacci, Navarro, Oman, Pakmor, Schaye, Simpson, \&
  Springel}]{Bose19}
Bose S. {et~al.}, 2019, Monthly Notices of the Royal Astronomical Society, 486,
  4790

\bibitem[{{Boylan-Kolchin}, {Bullock} \& {Kaplinghat}(2011){Boylan-Kolchin},
  {Bullock}, \& {Kaplinghat}}]{boylan11}
{Boylan-Kolchin} M., {Bullock} J.~S., {Kaplinghat} M., 2011, \mnras, 415, L40

\bibitem[{Bozek {et~al}\mbox{.}(2018)Bozek, Fitts, Boylan-Kolchin,
  Garrison-Kimmel, Abazajian, Bullock, Kere{\v{s}}, Faucher-Gigu{\`{e}}re,
  Wetzel, Feldmann, \& Hopkins}]{Bozek18}
Bozek B. {et~al.}, 2018, Monthly Notices of the Royal Astronomical Society,
  483, 4086

\bibitem[{Bullock \& Boylan-Kolchin(2017)}]{bullock17}
Bullock J.~S., Boylan-Kolchin M., 2017, Annual Review of Astronomy and
  Astrophysics, 55, 343

\bibitem[{Burger \& Zavala(2019)}]{burger19}
Burger J.~D., Zavala J., 2019, Monthly Notices of the Royal Astronomical
  Society, 485, 1008

\bibitem[{Chamberlain(2015)}]{ChapelChapterBalajiBook}
Chamberlain B.~L., 2015, in Programming Models for Parallel Computing, Balaji
  P., ed., MIT Press, pp. 129--159

\bibitem[{Chamberlain, Callahan \& Zima(2007)Chamberlain, Callahan, \&
  Zima}]{ChapelIJHPCA}
Chamberlain B.~L., Callahan D., Zima H.~P., 2007, International Journal of High
  Performance Computing Applications, 21, 291

\bibitem[{{Chan} {et~al}\mbox{.}(2022){Chan}, {Ferreira}, {May}, {Hayashi}, \&
  {Chiba}}]{2022MNRAS.511..943C}
{Chan} H. Y.~J., {Ferreira} E. G.~M., {May} S., {Hayashi} K., {Chiba} M., 2022,
  \mnras, 511, 943

\bibitem[{Chan {et~al}\mbox{.}(2022)Chan, Ferreira, May, Hayashi, \&
  Chiba}]{chan22}
Chan H. Y.~J., Ferreira E. G.~M., May S., Hayashi K., Chiba M., 2022, Monthly
  Notices of the Royal Astronomical Society, 511, 943

\bibitem[{Chan {et~al}\mbox{.}(2015)Chan, Kere\^s, O\~norbe, Hopkins, Muratov,
  Faucher-Gigu\`ere, \& Quataert}]{chan15}
Chan T.~K., Kere\^s D., O\~norbe J., Hopkins P.~F., Muratov A.~L.,
  Faucher-Gigu\`ere C.-A., Quataert E., 2015, Monthly Notices of the Royal
  Astronomical Society, 454, 2981

\bibitem[{Chen, Schive \& Chiueh(2017)Chen, Schive, \& Chiueh}]{chen17}
Chen S.-R., Schive H.-Y., Chiueh T., 2017, Monthly Notices of the Royal
  Astronomical Society, 468, 1338

\bibitem[{Chowdhury {et~al}\mbox{.}(2021)Chowdhury, van~den Bosch, Robles, van
  Dokkum, Schive, Chiueh, \& Broadhurst}]{dutta21}
Chowdhury D.~D., van~den Bosch F.~C., Robles V.~H., van Dokkum P., Schive
  H.-Y., Chiueh T., Broadhurst T., 2021, The Astrophysical Journal, 916, 27

\bibitem[{Collins {et~al}\mbox{.}(2013)Collins, Chapman, Rich, Ibata, Martin,
  Irwin, Bate, Lewis, Peñarrubia, Arimoto, Casey, Ferguson, Koch, McConnachie,
  \& Tanvir}]{collins13}
Collins M. L.~M. {et~al.}, 2013, The Astrophysical Journal, 768, 172

\bibitem[{Colín {et~al}\mbox{.}(2015)Colín, Avila-Reese, González-Samaniego,
  \& Velázquez}]{Colin15}
Colín P., Avila-Reese V., González-Samaniego A., Velázquez H., 2015, The
  Astrophysical Journal, 803, 28

\bibitem[{de~Blok {et~al}\mbox{.}(2008)de~Blok, Walter, Brinks, Trachternach,
  Oh, \& Kennicutt}]{deBlok08}
de~Blok W. J.~G., Walter F., Brinks E., Trachternach C., Oh S.-H., Kennicutt
  R.~C., 2008, The Astronomical Journal, 136, 2648

\bibitem[{de~Naray, McGaugh \& de~Blok(2008)de~Naray, McGaugh, \&
  de~Blok}]{deNaray08}
de~Naray R.~K., McGaugh S.~S., de~Blok W. J.~G., 2008, The Astrophysical
  Journal, 676, 920

\bibitem[{Di~Cintio {et~al}\mbox{.}(2014)Di~Cintio, Brook, Macci\`o, Stinson,
  Knebe, Dutton, \& Wadsley}]{dicintio14}
Di~Cintio A., Brook C.~B., Macci\`o A.~V., Stinson G.~S., Knebe A., Dutton
  A.~A., Wadsley J., 2014, Monthly Notices of the Royal Astronomical Society,
  437, 415

\bibitem[{Dutton {et~al}\mbox{.}(2020)Dutton, Buck, Macciò, Dixon, Blank, \&
  Obreja}]{Dutton20}
Dutton A.~A., Buck T., Macciò A.~V., Dixon K.~L., Blank M., Obreja A., 2020,
  Monthly Notices of the Royal Astronomical Society, 499, 2648

\bibitem[{Fitts {et~al}\mbox{.}(2019)Fitts, Boylan-Kolchin, Bozek, Bullock,
  Graus, Robles, Hopkins, El-Badry, Garrison-Kimmel, Faucher-Gigu{\`{e}}re,
  Wetzel, \& Kere{\v{s}}}]{Fitts19}
Fitts A. {et~al.}, 2019, Monthly Notices of the Royal Astronomical Society,
  490, 962

\bibitem[{Fitts {et~al}\mbox{.}(2017)Fitts, Boylan-Kolchin, Elbert, Bullock,
  Hopkins, O?orbe, Wetzel, Wheeler, Faucher-Gigu?re, Kere?, Skillman, \&
  Weisz}]{fitts17}
---, 2017, Monthly Notices of the Royal Astronomical Society, 471, 3547

\bibitem[{{Fraternali, F.} {et~al}\mbox{.}(2009){Fraternali, F.}, {Tolstoy,
  E.}, {Irwin, M. J.}, \& {Cole, A. A.}}]{fraternali09}
{Fraternali, F.}, {Tolstoy, E.}, {Irwin, M. J.}, {Cole, A. A.}, 2009, A\&A,
  499, 121

\bibitem[{Freundlich {et~al}\mbox{.}(2019)Freundlich, Dekel, Jiang, Ishai,
  Cornuault, Lapiner, Dutton, \& Macci{\`{o}}}]{Freundlich19}
Freundlich J., Dekel A., Jiang F., Ishai G., Cornuault N., Lapiner S., Dutton
  A.~A., Macci{\`{o}} A.~V., 2019, Monthly Notices of the Royal Astronomical
  Society, 491, 4523

\bibitem[{Fry {et~al}\mbox{.}(2015)Fry, Governato, Pontzen, Quinn, Tremmel,
  Anderson, Menon, Brooks, \& Wadsley}]{fry15}
Fry A.~B. {et~al.}, 2015, Monthly Notices of the Royal Astronomical Society,
  452, 1468

\bibitem[{Garrison-Kimmel {et~al}\mbox{.}(2019)Garrison-Kimmel, Hopkins,
  Wetzel, Bullock, Boylan-Kolchin, Kereš, Faucher-Giguère, El-Badry,
  Lamberts, Quataert, \& Sanderson}]{garrison19}
Garrison-Kimmel S. {et~al.}, 2019, Monthly Notices of the Royal Astronomical
  Society, 487, 1380

\bibitem[{Garrison-Kimmel {et~al}\mbox{.}(2013)Garrison-Kimmel, Rocha,
  Boylan-Kolchin, Bullock, \& Lally}]{shea13}
Garrison-Kimmel S., Rocha M., Boylan-Kolchin M., Bullock J.~S., Lally J., 2013,
  Monthly Notices of the Royal Astronomical Society, 433, 3539

\bibitem[{Genina {et~al}\mbox{.}(2017)Genina, Benítez-Llambay, Frenk, Cole,
  Fattahi, Navarro, Oman, Sawala, \& Theuns}]{genina17}
Genina A. {et~al.}, 2017, Monthly Notices of the Royal Astronomical Society,
  474, 1398

\bibitem[{Governato {et~al}\mbox{.}(2012)Governato, Zolotov, Pontzen,
  Christensen, Oh, Brooks, Quinn, Shen, \& Wadsley}]{governato12}
Governato F. {et~al.}, 2012, Monthly Notices of the Royal Astronomical Society,
  422, 1231

\bibitem[{Guzm\'an \& Ure\~na L\'opez(2004)}]{guzman04}
Guzm\'an F.~S., Ure\~na L\'opez L.~A., 2004, Phys. Rev. D, 69, 124033

\bibitem[{{Hernquist}(1990)}]{Hernquist90}
{Hernquist} L., 1990, \apj, 356, 359

\bibitem[{{Hu}, {Barkana} \& {Gruzinov}(2000){Hu}, {Barkana}, \&
  {Gruzinov}}]{hu00}
{Hu} W., {Barkana} R., {Gruzinov} A., 2000, Physical Review Letters, 85, 1158

\bibitem[{Hui {et~al}\mbox{.}(2017)Hui, Ostriker, Tremaine, \& Witten}]{hui17}
Hui L., Ostriker J.~P., Tremaine S., Witten E., 2017, Phys. Rev. D, 95, 043541

\bibitem[{Ir\ifmmode \check{s}\else \v{s}\fi{}i\ifmmode~\check{c}\else
  \v{c}\fi{} {et~al}\mbox{.}(2017)Ir\ifmmode \check{s}\else
  \v{s}\fi{}i\ifmmode~\check{c}\else \v{c}\fi{}, Viel, Haehnelt, Bolton, \&
  Becker}]{irsic17}
Ir\ifmmode \check{s}\else \v{s}\fi{}i\ifmmode~\check{c}\else \v{c}\fi{} V.,
  Viel M., Haehnelt M.~G., Bolton J.~S., Becker G.~D., 2017, Phys. Rev. Lett.,
  119, 031302

\bibitem[{Ji \& Sin(1994)}]{ji94}
Ji S.~U., Sin S.~J., 1994, Phys. Rev. D, 50, 3655

\bibitem[{Kendall \& Easther(2020)}]{kendall20}
Kendall E., Easther R., 2020, Publications of the Astronomical Society of
  Australia, 37, e009

\bibitem[{Kirby {et~al}\mbox{.}(2014)Kirby, Bullock, Boylan-Kolchin,
  Kaplinghat, \& Cohen}]{kirby14}
Kirby E.~N., Bullock J.~S., Boylan-Kolchin M., Kaplinghat M., Cohen J.~G.,
  2014, Monthly Notices of the Royal Astronomical Society, 439, 1015

\bibitem[{Lazar {et~al}\mbox{.}(2020)Lazar, Bullock, Boylan-Kolchin, Chan,
  Hopkins, Graus, Wetzel, El-Badry, Wheeler, Straight, Kereš,
  Faucher-Giguère, Fitts, \& Garrison-Kimmel}]{lazar20}
Lazar A. {et~al.}, 2020, Monthly Notices of the Royal Astronomical Society,
  497, 2393

\bibitem[{Leaman {et~al}\mbox{.}(2012)Leaman, Venn, Brooks, Battaglia, Cole,
  Ibata, Irwin, McConnachie, Mendel, \& Tolstoy}]{Leaman_2012}
Leaman R. {et~al.}, 2012, The Astrophysical Journal, 750, 33

\bibitem[{Lee \& Koh(1996)}]{lee96}
Lee J.-w., Koh I.-g., 1996, Phys. Rev. D, 53, 2236

\bibitem[{Lora(2015)}]{Lora15}
Lora V., 2015, The Astrophysical Journal, 807, 116

\bibitem[{Matos, Guzm\'an \& {Ure\~na-L\'opez}(2000)Matos, Guzm\'an, \&
  {Ure\~na-L\'opez}}]{matos00}
Matos T., Guzm\'an F., {Ure\~na-L\'opez} L.~A., 2000, Classical and Quantum
  Gravity, 17, 1707

\bibitem[{May \& Springel(2021)}]{May21}
May S., Springel V., 2021, Monthly Notices of the Royal Astronomical Society,
  506, 2603

\bibitem[{McConnachie(2012)}]{mcconnachie12}
McConnachie A.~W., 2012, The Astronomical Journal, 144, 4

\bibitem[{Mocz {et~al}\mbox{.}(2019)Mocz, Fialkov, Vogelsberger, Becerra, Amin,
  Bose, Boylan-Kolchin, Chavanis, Hernquist, Lancaster, Marinacci, Robles, \&
  Zavala}]{mocz19}
Mocz P. {et~al.}, 2019, Phys. Rev. Lett., 123, 141301

\bibitem[{Mocz {et~al}\mbox{.}(2020)Mocz, Fialkov, Vogelsberger, Becerra, Shen,
  Robles, Amin, Zavala, Boylan-Kolchin, Bose, Marinacci, Chavanis, Lancaster,
  \& Hernquist}]{mocz20}
---, 2020, Monthly Notices of the Royal Astronomical Society, 494, 2027

\bibitem[{Mocz {et~al}\mbox{.}(2017)Mocz, Vogelsberger, Robles, Zavala,
  Boylan-Kolchin, Fialkov, \& Hernquist}]{mocz17}
Mocz P., Vogelsberger M., Robles V.~H., Zavala J., Boylan-Kolchin M., Fialkov
  A., Hernquist L., 2017, Monthly Notices of the Royal Astronomical Society,
  471, 4559–4570

\bibitem[{{Moore}(1994)}]{moore1994}
{Moore} B., 1994, \nat, 370, 629

\bibitem[{{Navarro}, {Eke} \& {Frenk}(1996){Navarro}, {Eke}, \&
  {Frenk}}]{navarro96}
{Navarro} J.~F., {Eke} V.~R., {Frenk} C.~S., 1996, \mnras, 283, L72

\bibitem[{Nori {et~al}\mbox{.}(2018)Nori, Murgia, Iršič, Baldi, \&
  Viel}]{Nori18}
Nori M., Murgia R., Iršič V., Baldi M., Viel M., 2018, Monthly Notices of the
  Royal Astronomical Society, 482, 3227

\bibitem[{Oman {et~al}\mbox{.}(2018)Oman, Marasco, Navarro, Frenk, Schaye, \&
  Benítez-Llambay}]{oman19}
Oman K.~A., Marasco A., Navarro J.~F., Frenk C.~S., Schaye J., Benítez-Llambay
  A., 2018, Monthly Notices of the Royal Astronomical Society, 482, 821

\bibitem[{{Padmanabhan} {et~al}\mbox{.}(2020){Padmanabhan}, {Ronaghan},
  {Zagorac}, \& {Easther}}]{chplUltraPaper}
{Padmanabhan} N., {Ronaghan} E., {Zagorac} J.~L., {Easther} R., 2020, in 2020
  IEEE International Parallel and Distributed Processing Symposium Workshops
  (IPDPSW), pp. 678--678

\bibitem[{Robles, Bullock \& Boylan-Kolchin(2018)Robles, Bullock, \&
  Boylan-Kolchin}]{rob18}
Robles V.~H., Bullock J.~S., Boylan-Kolchin M., 2018, Monthly Notices of the
  Royal Astronomical Society, 483, 289

\bibitem[{Robles {et~al}\mbox{.}(2017)Robles, Bullock, Elbert, Fitts,
  González-Samaniego, Boylan-Kolchin, Hopkins, Faucher-Giguère, Kereš, \&
  Hayward}]{rob17}
Robles V.~H. {et~al.}, 2017, Monthly Notices of the Royal Astronomical Society,
  472, 2945

\bibitem[{Robles \& Matos(2012)}]{rob12}
Robles V.~H., Matos T., 2012, \mnras, 422, 282

\bibitem[{Santos-Santos {et~al}\mbox{.}(2020)Santos-Santos, Navarro, Robertson,
  Benítez-Llambay, Oman, Lovell, Frenk, Ludlow, Fattahi, \& Ritz}]{santos20}
Santos-Santos I. M.~E. {et~al.}, 2020, Monthly Notices of the Royal
  Astronomical Society, 495, 58

\bibitem[{{Schive}, {Chiueh} \& {Broadhurst}(2014){Schive}, {Chiueh}, \&
  {Broadhurst}}]{sch14}
{Schive} H.-Y., {Chiueh} T., {Broadhurst} T., 2014, Nature Physics, 10, 496

\bibitem[{Schive {et~al}\mbox{.}(2014)Schive, Liao, Woo, Wong, Chiueh,
  Broadhurst, \& Hwang}]{sch14b}
Schive H.-Y., Liao M.-H., Woo T.-P., Wong S.-K., Chiueh T., Broadhurst T.,
  Hwang W.-Y.~P., 2014, Phys. Rev. Lett., 113, 261302

\bibitem[{Simon \& Geha(2007)}]{simon07}
Simon J.~D., Geha M., 2007, The Astrophysical Journal, 670, 313

\bibitem[{Strigari {et~al}\mbox{.}(2008)Strigari, Bullock, Kaplinghat, Simon,
  Geha, Willman, \& Walker}]{Strigari08}
Strigari L.~E., Bullock J.~S., Kaplinghat M., Simon J.~D., Geha M., Willman B.,
  Walker M.~G., 2008, Nature, 454, 1096

\bibitem[{{Su{\'a}rez}, {Robles} \& {Matos}(2014){Su{\'a}rez}, {Robles}, \&
  {Matos}}]{suarez14}
{Su{\'a}rez} A., {Robles} V.~H., {Matos} T., 2014, in Astrophysics and Space
  Science Proceedings, Vol.~38, Accelerated Cosmic Expansion, {Moreno
  Gonz{\'a}lez} C., {Madriz Aguilar} J.~E., {Reyes Barrera} L.~M., eds., p. 107

\bibitem[{Tollet {et~al}\mbox{.}(2016)Tollet, Macci\`{o}, Dutton, Stinson,
  Wang, Penzo, Gutcke, Buck, Kang, Brook, Di~Cintio, Keller, \&
  Wadsley}]{tollet16}
Tollet E. {et~al.}, 2016, Monthly Notices of the Royal Astronomical Society,
  456, 3542

\bibitem[{Ureña-López(2019)}]{urena19}
Ureña-López L.~A., 2019, Frontiers in Astronomy and Space Sciences, 6

\bibitem[{Veltmaat, Niemeyer \& Schwabe(2018)Veltmaat, Niemeyer, \&
  Schwabe}]{veltmaat18}
Veltmaat J., Niemeyer J.~C., Schwabe B., 2018, Phys. Rev. D, 98, 043509

\bibitem[{Veltmaat, Schwabe \& Niemeyer(2020)Veltmaat, Schwabe, \&
  Niemeyer}]{Veltmaat20}
Veltmaat J., Schwabe B., Niemeyer J.~C., 2020, Physical Review D, 101

\bibitem[{Vogelsberger {et~al}\mbox{.}(2014)Vogelsberger, Zavala, Simpson, \&
  Jenkins}]{Vogelsberger14}
Vogelsberger M., Zavala J., Simpson C., Jenkins A., 2014, Monthly Notices of
  the Royal Astronomical Society, 444, 3684

\bibitem[{{Walker} \& {Pe{\~n}arrubia}(2011)}]{wp11}
{Walker} M.~G., {Pe{\~n}arrubia} J., 2011, \apj, 742, 20

\bibitem[{Wheeler {et~al}\mbox{.}(2019)Wheeler, Hopkins, Pace, Garrison-Kimmel,
  Boylan-Kolchin, Wetzel, Bullock, Kereš, Faucher-Giguère, \&
  Quataert}]{wheeler19}
Wheeler C. {et~al.}, 2019, Monthly Notices of the Royal Astronomical Society,
  490, 4447

\bibitem[{Wolf {et~al}\mbox{.}(2010)Wolf, Martinez, Bullock, Kaplinghat, Geha,
  Mu\~{n}oz, Simon, \& Avedo}]{wolf10}
Wolf J., Martinez G.~D., Bullock J.~S., Kaplinghat M., Geha M., Mu\~{n}oz
  R.~R., Simon J.~D., Avedo F.~F., 2010, Monthly Notices of the Royal
  Astronomical Society, 406, 1220

\bibitem[{{Yavetz}, {Li} \& {Hui}(2022){Yavetz}, {Li}, \&
  {Hui}}]{2022PhRvD.105b3512Y}
{Yavetz} T.~D., {Li} X., {Hui} L., 2022, \prd, 105, 023512

\bibitem[{{Zagorac} {et~al}\mbox{.}(2023){Zagorac}, {Kendall}, {Padmanabhan},
  \& {Easther}}]{2023PhRvD.107h3513Z}
{Zagorac} J.~L., {Kendall} E., {Padmanabhan} N., {Easther} R., 2023, \prd, 107,
  083513

\bibitem[{{Zagorac} {et~al}\mbox{.}(2022){Zagorac}, {Sands}, {Padmanabhan}, \&
  {Easther}}]{2022PhRvD.105j3506Z}
{Zagorac} J.~L., {Sands} I., {Padmanabhan} N., {Easther} R., 2022, \prd, 105,
  103506

\bibitem[{Zhang {et~al}\mbox{.}(2018)Zhang, Kuo, Liu, Tsai, Cheung, \&
  Chu}]{Zhang18}
Zhang J., Kuo J.-L., Liu H., Tsai Y.-L.~S., Cheung K., Chu M.-C., 2018, The
  Astrophysical Journal, 863, 73

\end{thebibliography}

\bsp
\label{lastpage}
\end{document}